\documentclass[12pt]{article}
 
\usepackage{amssymb}     
\usepackage{latexsym}

\newcommand{\bq}{\begin{quote}}
\newcommand{\eq}{\end{quote}}

\newtheorem{Th}{Theorem}  
\newtheorem{ax}{Axiom}  
\newtheorem{lm}{Lemma} 
\newtheorem{df}{Definition}    
\newtheorem{pr}{Proposition} 
\newtheorem{cl}{Corollary}  
\newtheorem{re}{Remark}    
\newtheorem{as}{Assumption}  
\newtheorem{wg}{Wild Guess}
\newtheorem{ex}{Example}
\newcommand{\bth}{\begin{Th}\hspace{-5pt}{\bf .} \ } 
\newcommand{\Eth}{\end{Th}}
\newcommand{\bax}{\begin{ax}\hspace{-5pt}{\bf .} \ } 
\newcommand{\eax}{\end{ax}}
\newcommand{\blm}{\begin{lm}\hspace{-5pt}{\bf .} \ }
\newcommand{\elm}{\end{lm}}
\newcommand{\bdf}{\begin{df}\hspace{-5pt}{\bf .} \ }   
\newcommand{\edf}{\end{df}} 
\newcommand{\bpr}{\begin{pr}\hspace{-5pt}{\bf .} \ } 
\newcommand{\epr}{\end{pr}}
\newcommand{\bcl}{\begin{cl}\hspace{-5pt}{\bf .} \ } 
\newcommand{\ecl}{\end{cl}}
\newcommand{\bre}{\begin{re}\hspace{-5pt}{\bf .} \ }
\newcommand{\ere}{\end{re}}
\newcommand{\bas}{\begin{as}\hspace{-5pt}{\bf .} \ }
\newcommand{\eas}{\end{as}}
\newcommand{\bwg}{\begin{wg}\hspace{-5pt}{\bf .} \ }
\newcommand{\ewg}{\end{wg}}
\newcommand{\bex}{\begin{ex}\hspace{-5pt}{\bf .} \ }  
\newcommand{\eex}{\end{ex}}
\newcommand{\bpf}{\noindent {\bf Proof:} }
\newcommand{\epf}{$\bullet$\par\vspace{1.8mm}\noindent}
\newcommand{\bit}{\begin{itemize}}
\newcommand{\eit}{\end{itemize}\par\noindent}
\newcommand{\ben}{\begin{enumerate}}
\newcommand{\een}{\end{enumerate}\par\noindent}
\newcommand{\beq}{\begin{equation}}
\newcommand{\eeq}{\end{equation}\par\noindent}
\newcommand{\beqa}{\begin{eqnarray*}}
\newcommand{\eeqa}{\end{eqnarray*}\par\noindent}
\newcommand{\beqn}{\begin{eqnarray}}  
\newcommand{\eeqn}{\end{eqnarray}\par\noindent}

\ifx\optionkeymacros\undefined\else\endinput\fi
\catcode`\=\active\def{\c c}        % option c
\catcode`\²=\active\def²{\leq}        % option , (math mode)
\catcode`\³=\active\def³{\geq}        % option . (math mode)
\catcode`\Ò=\active\defÒ{``}          % option [
\catcode`\Ô=\active\defÔ{`}           % option ] 
\catcode`\Õ=\active\defÕ{'}           % shift option ]
\catcode`\'=\active\def'{\c C}        % shift-option C
\catcode`\Î=\active\defÎ{{\OE}}       % shift-option Q
\catcode`\¡=\active\def¡{\accent'27}  % shift-option 8
\catcode`\Ó=\active\defÓ{''}          % shift-option [
\catcode`\Š=\active\defŠ{\"a}        % option u, then  a
\catcode`\'=\active\def'{\"e}        % option u, then  e
\catcode`\•=\active\def•{\"{\i}}     % option u, then  i
\catcode`\š=\active\defš{\"o}        % option u, then  o
\catcode`\Ÿ=\active\defŸ{\"u}        % option u, then  u
\catcode`\Ø=\active\defØ{\"y}        % option u, then  y
\catcode`\€=\active\def€{\"A}        % option u, then  A
\catcode`\…=\active\def…{\"O}        % option u, then  O
\catcode`\†=\active\def†{\"U}        % option u, then  U
\catcode`\‡=\active\def‡{\'a}        % option e, then  a
\catcode`\Ž=\active\defŽ{\'e}        % option e, then  e
\catcode`\'=\active\def'{\'{\i}}     % option e, then  i
\catcode`\—=\active\def—{\'o}        % option e, then  o
\catcode`\œ=\active\defœ{\'u}        % option e, then  u
\catcode`\ƒ=\active\defƒ{\'E}        % option e, then  E
\catcode`\ˆ=\active\defˆ{\`a}        % option `, then  a
\catcode`\=\active\def{\`e}        % option `, then  e
\catcode`\"=\active\def"{\`{\i}}     % option `, then  i
\catcode`\˜=\active\def˜{\`o}        % option `, then  o
\catcode`\=\active\def{\`u}        % option `, then  u
\catcode`\Ë=\active\defË{\`A}        % option `, then  A
\catcode`\‹=\active\def‹{\~a}        % option n, then  a
\catcode`\–=\active\def–{\~n}        % option n, then  n
\catcode`\›=\active\def›{\~o}        % option n, then  o
\catcode`\Ì=\active\defÌ{\~A}        % option n, then  A
\catcode`\"=\active\def"{\~N}        % option n, then  N
\catcode`\Í=\active\defÍ{\~O}        % option n, then  O
\catcode`\‰=\active\def‰{\^a}        % option i, then  a
\catcode`\=\active\def{\^e}        % option i, then  e
\catcode`\"=\active\def"{\^{\i}}     % option i, then  i
\catcode`\™=\active\def™{\^o}        % option i, then  o
\catcode`\ž=\active\defž{\^u}        % option i, then  u 
\catcode`\¾=\active\def¾{{\ae}}      % option '
\let\optionkeymacros\null
\begin{document}   
\noindent\centerline{\large{\bf The Sasaki Hook is not a [Static] Implicative Connective}}
 
\smallskip\noindent\centerline{\large{\bf but Induces a Backward 
[in Time] Dynamic One that Assigns Causes}}  

\smallskip\par\noindent
\smallskip\par\noindent
\centerline{\normalsize{Bob Coecke}}
\par\smallskip\noindent 
\centerline{\footnotesize{University of Cambridge, Department of Pure Mathematics and Mathematical
Statistics\,, and,}}
\par\smallskip\noindent 
\centerline{\footnotesize{University of Oxford, Oxford University Computing Laboratory (preferred mail
address)\,,}}\vspace{-1mm}
\par\noindent 
\centerline{\footnotesize{Wolfson Building, Parks Road, Oxford, OX1 3QD, UK\,; e-mail: coecke@comlab.ox.ac.uk\,.}}  
\par\medskip\par\smallskip\noindent 
\centerline{{Sonja Smets}}
\par\smallskip\noindent
\centerline{\footnotesize{Free University of
Brussels (VUB), Department of Philosophy,}}\vspace{-1mm} 
\par\noindent 
\centerline{\footnotesize{Pleinlaan 2, B-1050
Brussels, Belgium\,; e-mail: sonsmets@vub.ac.be\,.}}  

\def\retro{\hbox{$\small\, \circ\hspace{-1mm} -\hspace{-1mm}\ $}}  

\par\medskip\noindent 
\begin{abstract}   
\par\noindent 
In this paper we argue that the Sasaki adjunction, which formally encodes the logicality
that different authors tried to attach to the Sasaki hook as a `quantum implicative
connective', has a fundamental dynamic nature and encodes the so-called `causal duality' (Coecke, Moore and Stubbe
2001) for the particular case of 
a quantum measurement with a projector as corresponding self-adjoint operator. 
In particular: The action of the Sasaki hook $(a\stackrel{S}{\to}-)$ for fixed antecedent $a$ assigns to some property
``the weakest cause before the measurement of actuality of that property after the measurement'', i.e.
${(a\stackrel{S}{\to}b)}$ is the weakest property that guarantees actuality of $b$ after performing the measurement
represented by the projector that has the `subspace $a$' as eigenstates for eigenvalue $1$\,, say, the measurement that
`tests' $a$\,.   
From this we conclude that the logicality attributable to quantum systems contains a fundamentally
dynamic ingredient: Causal duality actually provides a new dynamic interpretation of orthomodularity.  
We also reconsider the status of the Sasaki hook within `dynamic (operational) quantum logic'
(DOQL), what leads us to the claim made in the title of this paper. 
More explicitly, although (as many argued in the past) the Sasaki hook should not be seen as an implicative hook, the
formal motivation that persuaded others to do so, i.e. the Sasaki adjunction, does have a physical significance (in
terms of causal duality). It is within the context of DOQL that we can then derive that the
labeled dynamic hooks (forwardly and backwardly) that encode
quantum measurements act on properties as
$(a_1\stackrel{\varphi_a}{\to}a_2):=
(a_1\to_L(a\stackrel{S}{\to} a_2))$ and
$(a_1\stackrel{\varphi_a}{\leftarrow}a_2):=
((a\stackrel{S}{\to}a_2)\to_L a_1)$\,, taking values in the
`disjunctive extension' ${\rm DI}(L)$ of the property lattice
$L$ (Coecke 2001a)\,, where $a\in L$ is the 
tested   
property and $(-\to_L-)$ is the Heyting implication
that lives on ${\rm DI}(L)$\,. Since these hooks $(-\stackrel{\varphi_a}{\to}-)$ and 
$(-\stackrel{\varphi_a}{\leftarrow}-)$ extend to ${\rm DI}(L)\times{\rm DI}(L)$ they constitute
internal operations\,.
In an even more radical perspective one could say that the transition from either classical or 
constructive/intuitionistic logic to quantum logic entails besides the introduction of an
additional unary connective `operational resolution' (Coecke 2001a) the shift from a binary connective implication to a
ternary connective where two of the arguments refer to qualities of the system and the third, the new one, to
an obtained outcome (in a measurement).
\end{abstract} 
 
\medskip\noindent
{\bf 1. QUANTUM LOGICALITY} 
     
\medskip\noindent
We claim that logical considerations on quantum behavior  
%%% 
%% 
%
and as such,  further development of the research field, 
% 
%%  
%%%
have been `corrupted' by two features.  Once these two features are neutralized,
the way towards an essentially dynamic quantum logic (i.e. a unified
logic of `changes' both for classical and quantum systems), or otherwise put,
a true quantum process semantics, is opened.  Moreover, the solution
to the second `corrupt feature' indicates that the logicality encoded
in pure quantum theory is of a fundamental dynamic nature. 
%%% 
%%  
%
Structures somewhat similar to those emerging in the context of categorical grammar (Lambek 1958), linear logic (Girard
1987, 2000), action logic (Baltag 1999) and computation and concurrency (Abramsky 1993; Milner 1999) then naturally emerge
via a Kripke style approach for logical semantics applied to the operational foundations of physics.    
%
%%
%%%
The two features that obstructed true logicality are --- concerning the
second, most non-quantum logicians have always agreed on its weakness:
\par\smallskip\noindent
i. The Birkhoff  and von Neumann (1936) `dilemma': ``Whereas logicians  have
usually assumed that [the orthocomplementation] properties L71-L73 of
negation were the ones least able to withstand a critical analysis, the
study of mechanics points to the distributive identities L6 as the
weakest link in the algebra of logic.''\ \ This dilemma forced the search and attempted
identification of quantum logicality to proceed `orthogonal' to intuitionistic and
derived developments in logic.
\par\smallskip\noindent
ii. `Implication' via the  Sasaki adjunction: The fact that the 
pointwise action $\varphi_{a}^*(-)$ of Hilbert space projectors on the subspace lattice, the quantum analogue of
the action of classical lattice projections $(a\wedge -)$\,, has the parameterized action $(a\stackrel{S}{\to} -)$
of the so-called Sasaki hook $(-\stackrel{S}{\to} -)$ as a right adjoint.
This Sasaki hook (as a binary operation) satisfies the minimal implicative condition $(a\stackrel{S}{\to}
b)=1\Longleftrightarrow a\leq b\,$ (Kalmbach 1983) where
$\leq$ naturally encodes physical consequence (Coecke, Moore and Smets 2001a).\footnote{Recall
that the adjointness of projection $(a\wedge -)$ and implication $(a\to-)$ in classical
and intuitionistic logic exactly encodes the validity of modus ponens and
deduction, in other words, the adjunction (sometimes called
the `implicative condition') $a\wedge x\leq b\ \Longleftrightarrow\ x\leq (a \to b)$ is equivalent to 
$a \wedge x \leq b\ \Longrightarrow\ x \leq (a \to b)$ together with $a \wedge (a \to b) \leq b$\,. By means of
applying the latter, i.e. modus ponens, given that $x \leq (a \to b)$ we indeed obtain $a \wedge x \leq a \wedge (a
\to b) \leq b$\,.  We come back to this further in this paper.} 
However, any proof theoretic consideration (among other things) did turn out to be impossible for a logical system
with $(-\stackrel{S}{\to}-)$ as implication since there cannot be a deduction theorem for it (Blok, K\"oler
and Pigozzi 1984; Malinowski 1990)\,.\footnote{In Hardegree (1975, 1979) and Herman, Marsden and Piziak
(1975) it is pointed out that the Sasaki hook also fails to satisfy
strong transitivity, weakening and contraposition.}  We explain all this in more detail in section 3. 
 
\smallskip
It turns out that an operational analysis of quantum logicality starting
from well-defined primitive notions rather than from formal pragmatism
eliminates these two features. Instead:
\par\smallskip\noindent
i. As shown in Coecke (2001a), the injective hull construction
for meet-semilattices (Bruns and Lakser 1970; Horn and Kimura 1971)  realizes a disjunctive
extension of property lattices, the latter being the physical incarnation of meet-complete and conjunctive quantum
logicality (and nothing more!),\footnote{For a demonstration of complete conjunctivity see Piron (1976) and Moore
(1999).  For the `and nothing more!' see Emch and Jauch (1965), Coecke (2001a) and Coecke, Moore and Smets (2001a).} in
terms of a complete Heyting algebra that goes equipped with an additional operation, `operational resolution', which
recaptures the initial property lattice as its range, and this goes without any loss of the (physically derivable)
logical content of the initial lattice of properties. In the case of an atomistic property lattice the inclusion of the
property lattice in its distributive hull actually encodes the `state space - property lattice duality' (Coecke
2001b).\footnote{For a discussion of the categorical `state space - property lattice
duality' for atomistic orthocomplemented lattices, physically and mathematically, see Moore (1995).}   
This construction will be recalled in the fourth section of this paper.
\par\smallskip\noindent
ii. Propagation of physical properties is left adjoint to backward   
causal assignment (Coecke, Moore and Stubbe 2001) --- we provide a more intuitive presentation of this result in
the section 5.    Also in section 5\,, we show that the Sasaki adjunction exactly encodes this
adjunction for the case of propagation and backward causal assignment of a
quantum measurement.   The minimal implicative condition expresses in this perspective merely that the image under
projection  of the trivial property $1$ is exactly the property on which we project, i.e. 
%%% 
%% 
%
$\varphi_{a}^*(1)=a$. 
More important however, recalling that adjointness of Sasaki hook and Sasaki projection for an ortholattice is equivalent
to the ortholattice being orthomodular, causal duality provides actually a new interpretation of orthomodularity.
% 
%%     
%%%  

\smallskip 
Moreover, as shown in Coecke (2001b) and Coecke, Moore and Smets (2001b), when combining
the following features that result from the above:
\par\smallskip\noindent
i. property lattices admit a canonical disjunctive extension giving rise to an
irredundant collection of meaningful propositions on properties with a physically
significant ordering,\footnote{For a clear distinction between the significance of `properties' and `propositions on
properties' we initially refer to Coecke (2001a) and the rest of this paper.  Briefly, from a philosophical perspective
one could say that properties are ontological there where propositions on properties are to be situated at an
epistemological meta-level.}
\par\smallskip\noindent
ii. a unary connective `operational resolution' faithfully
recaptures the physical properties within the collection of propositions on these properties as its range, and,  
\par\smallskip\noindent 
%%% 
%% 
%
iii. causal duality applies both to properties and to propositions on
properties, respectively restricting physically admissible evolution, and encoding preservation of
propositional disjunction. 
% 
%%     
%%%
\par\smallskip\noindent 
then, a Kripke-style approach for logical semantics applied to the operational foundations of physics yields a
logical structure with for each possible physical `environment' (e.g., a measurement apparatus, a free or imposed
evolution, interaction in the presence of another system, etc.) the following connectives:
\par\smallskip\noindent
i. two implications $(-\stackrel{e}{\to}-)$ and $(-\stackrel{e}{\leftarrow}-)$ that extend the
physical content of propagation of (physical) properties and backward causal assignment, and, 
\par\smallskip\noindent
ii. two corresponding adjoint tensors $(-\otimes_e-)$ and $(-{_e \otimes}-)$ of which one is
commutative and one isn't.  
 
\smallskip\noindent
This, since the Sasaki adjunction encodes causal duality, then establishes our claim made in the title
concerning the induced dynamic implications by the Sasaki hook.  In the `static limit', i.e. `freezed dynamics' with
respect to some preferred referential frame for space-like properties, this structure yields an intuitionistic logic
equipped with the above mentioned operational resolution as an additional operation, and both the hooks
$(-\stackrel{e}{\to}-)$ and
$(-\stackrel{e}{\leftarrow}-)$ collapse into the [static] Heyting implication, and the tensors
$(-\otimes_e-)$ and $(-{_e\otimes}-)$ become binary conjunction.\footnote{The multiplicative fragments respectively provide
a commutative quantale and dual non-commutative quantale semantics (Coecke 2001b; Coecke, Moore and Smets 2001b; Smets
2001)\,.}   We also recall here the following spin-off from all the
above (for details we refer to corresponding cited papers):    
\par\smallskip\noindent
i. A proof of linearity for deterministic evolution and for the Hilbert space tensor product
as a description of quantum compoundness (Faure, Moore and Piron 1995; Coecke 2000).
\par\smallskip\noindent
ii. A generalized notion of linearity for indeterministic transitions that
saturates into ordinary linearity in the deterministic case (Coecke and Stubbe 1999; Coecke,
Stubbe and Valckenborgh 2001).
\par\smallskip\noindent
iii. A counter example to van Benthem's (1991, 1994) `general dynamic logic in terms
of relational structures': relational inverses have not necessarily any physical significance
for non-classical systems (Coecke, Moore and Smets 2001b).   
  
\smallskip
We will proceed as follows in this paper: Since we feel very strong about the fact that quantum
logicality cannot be treated as a purely mathematical matter without specifying what one is
actually talking about and that in every other case it might even be better to abandon the word
quantum (at least as a reference to physics) in ones discourse, we provide in the next section
an outline of the primitive physical notions from which we derive our
formal notions.\footnote{Obviously there is
something to say for the use of the word quantum referring to a domain of mathematics that studies structures
inspired on particular formal features of the quantum mechanical formalism such as non-distributivity, but this
still remains pure mathematics in absence of an outline of the primitive physical notions from which one derives
formal notions such as order, bounds and in particular of the significance of elements in any considered set on
which one defines these relations and connectives.  In this context,
for a recent survey of general operational quantum logic we refer to Coecke, Moore and Wilce (2000).} Next, we recall some mathematical preliminaries required for this paper 
including Galois adjoints, Heyting
algebras and the Sasaki adjunction itself. In
section 4\,, besides briefly recalling the results in Coecke (2001a), we discuss the Sasaki hook in perspective of
these, in particular we argue that any true implicative connective on the lattice of properties of a quantum system
has to be external and as such cannot be the Sasaki hook. 
In the fifth section, besides explaining causal duality and as such the true significance of the Sasaki adjunction,
we introduce `dynamic causal relations' which express the intuitive contents of the Sasaki adjunction in an
alternative way.  These relations form the core of our approach in the sixth section where the formal content of the
Sasaki adjunction will be implemented in the
framework of DOQL --- see also Coecke (2001b)\,, Coecke, Moore and
Smets (2001b) and Smets (2001)\,. Our analysis in this paper
ends with an overview of the dynamic implications $(-\stackrel{\varphi_a}{\to}-)$ and 
$(-\stackrel{\varphi_a}{\leftarrow}-)$ which we can deduce
from the Sasaki adjunction.  Finally, section 7 points to the possible impact of our
approach on the field of quantum logic and opens new perspectives to be elaborated on in the
future.

\bigskip\noindent
{\bf 2. WHAT QUANTUM LOGICALITY CAN BE ABOUT}     
     
\medskip\noindent
We claim that it makes no sense to discuss quantum logicality without specifying what the elements in
the considered lattice physically stand for.  Indeed, nonsense arguments, as for
example indicated at in Foulis and Randall (1984) and Piziak (1986), emerge due to conceptual
mixup.\footnote{We rather not refer to the papers containing mathematical/conceptual flaws but give credit to
those who tackled them.}   See also Smets (2001 \S 6) for a more general survey on misunderstandings and
misconceptions on physical logicality.  To situate our perspective clearly we will recall here two major
(well-defined) perspectives which are, although essentially different respectively being ontological (Jauch and
Piron 1969; Piron 1976) and empirical (Foulis and Randall 1972; Randall and Foulis 1973), not at all exclusive
(Foulis, Piron and Randall 1983), but which give rise to different mathematical structures --- see for example
Coecke, Moore and Wilce (2000) for an overview and Moore (1999) and Wilce (2000) for recent surveys respectively on
the Jauch-Piron and the Foulis-Randall perspective.   How can one theoretically approach the behavior of a physical
system?  As philosophers know very well (to whom physicist however in general don't pay much
attention)\,\footnote{See for example Rovelli (1999) who backs us up on this: ``I am convinced of the reciprocal
usefulness of a dialog between  physics and philosophy (Rovelli 1997).  This dialog has played a  major role during
the other periods in which science faced  foundational problems.  In my opinion, most physicists  underestimate the
effect of their own epistemological prejudices  on their research [...] On the one hand, a more acute philosophical
awareness  would greatly help the physicists engaged in fundamental 
research: Newton, Heisenberg and Einstein couldn't have done what      
they have done if they weren't nurtured by (good or bad) 
philosophy.''} there are different answers to this question.  As such, any approach
requires a subtle specification of what the primitive notions are one starts from. In   
Foulis and Randall (1972) and Randall and Foulis (1973) one considers a notion to which we 
prefer to refer to as ``observed events that reflect something about the system's
qualities'', where in Jauch and Piron (1969) and Piron (1976) one considers ``qualities of
the system that cause certain events to occur'', depending on the particular environment (e.g., presence of a
measurement device). As we know from quantum mechanics, the state of the system in general doesn't determine the
outcome of a measurement, and, an event provoked by a measurement actually changes the system's qualities. As such, it
comes as no surprise that these perspectives yield different mathematical structures.
To a certain extend one could say that both in the Jauch-Piron and Foulis-Randall perspective we are
interested in how the system interacts with its environment, although in the first case from the `system's
perspective' where in the second case we rather consider the `environments perspective', including the physicist
that effectuates the experiments, or in other words, an endo- versus an exo-perspective --- see also Coecke
(2001b) for a discussion on this matter, slightly deviating from the original Jauch-Piron approach allowing some
additional flexibility in view of actual applications.  Obviously, since the Foulis-Randall perspective is an
exo-perspective, the measurements are made explicit within the formalism.  Their formalism is indeed
essentially about how the system's behavior is reflected through measurements, without specifying the behavior itself. 
In the Jauch-Piron perspective where we focus on the system's behavior itself this is a somewhat more subtle matter. 
Since it adopts an endo-perspective, the measurement is not a priori part of the `universe of discourse'.  Therefore it
will be incorporated in a conditional way, explicitly as ``a system in a particular realization $p$, i.e. state,
possesses a quality $a$ if it is the case that:  whenever it (in realization
$p$) is within environment
$e_a$ then it causes phenomenon
$\alpha_a$ to happen'' and it is by this statement that we identify a particular quality of the system\footnote{Note
here also that ``whenever the system is within environment $e_a$ then it causes phenomenon
$\alpha_a$ to happen'' corresponds with Piron's ``whenever a definite
experimental project is effectuated we obtain a positive outcome with certainty'' (Piron 1976; Moore 1999) where the
definite experimental project includes both a physical procedure, say placing the system within environment $e_a$, and
specification of what is a positive answer to this procedure, say phenomenon $\alpha_a$ happens.  By referring to a
causal connection we aim to avoid the confusion raised by use of the notion `certainty' in Piron's formulation. One
could also more naively say that Piron's formulation is an active one (from a physicist's perspective) where our's is a
passive one.  Again, by the passive formulation we avoid any connotation with some role that is in many interpretations
of quantum theory  ascribed to the so-called `observer'.} --- this explicit consideration of the environment
(or context), even in the system's endo-perspective, is what gives the operational flavour to this 
approach.\footnote{Note
that operationalism has here nothing to do with instrumentalism.  In
Piron's formulation the  tendency towards an instrumentalist
interpretation is however a bit stronger due to the explicit presence of `definite
experimental projects'.  By considering general environments instead of
specific physical procedures we hope to avoid some confusion and eliminate the link to P.W.
Bridgman's operationalism since in our case, physical qualities have an
extension in reality and are not by means of definitions reducible to sets
of procedures --- see Smets (2001 \S 1).} 

\smallskip
Before we continue let us first recall some basic order theoretical notions.  A {\sl complete lattice} is a bounded 
partially ordered set
$(L,\leq, 0, 1)$ which is such that every subset
$A\subseteq L$ has a greatest lower bound or meet $\bigwedge A$. It then follows that every subset
$A\subseteq L$ also has a smallest upper bound or join $\bigvee A$ via {\sl Birkhoff's theorem}: 
\beq\label{eq:birk}  
\bigvee A=\bigwedge\{b\in L|\,\forall a\in A:b\geq a\}\,.  
\eeq  
If the bounded poset $(L,\leq, 0, 1)$ only admits finite greatest lower bounds and finite least upper bounds we
call it a {\sl lattice}.  In case it has only finite greatest lower bounds and not necessarily least upper
bounds we call it a {\sl meet-semilattice}.   A first main example of a complete lattice is the lattice $L_{\cal H}$ of
closed subspaces of a Hilbert space ${\cal H}$, ordered by inclusion, or, isomorphically, the lattice of
orthogonal projectors on this Hilbert space, ordered via $P_A\leq P_B\Leftrightarrow P_B\circ 
P_A=P_A\circ P_B=P_A$\,, i.e. if and only if we have
$A\subseteq B$ for the corresponding subspaces (Dunford and Schwartz 1957 \S VI.3).  In the closed
subspace perspective meets correspond to intersection and joins to closed linear span. In the  
projector perspective it is harder to grasp the operations meets and join since they only can be
expressed in a simple tangible way in case of commuting projectors (Dunford
and Schwartz 1957 \S VI.3). A second example is the powerset ${\rm P}(X)$ of any set
$X$\,, i.e. the set of subsets of this set, ordered again by inclusion and meets and joins are
respectively intersection and union.  Orthomodular lattices generalize these two cases of the
Hilbert space projection lattice and the powerset of a set.  Recall here that an {\sl orthomodular lattice}
is a lattice that goes equipped with an {\sl orthocomplementation} $':{L}\to{L}$, defined by
$a\leq b\Rightarrow b'\leq a'$,  $a\wedge a'=0$, $a\vee a'=1$ and $a''=a$, 
and which is such that  $a\leq b$ implies $a\vee(a'\wedge b)=b$\,.    Alternative characterizations of
orthomodularity can be found in section 3 of this paper. One verifies that every modular ortholattice is also an
orthomodular lattice, and for that reason one refers to the additional property an orthomodular lattice has compared to
an ortholattice as weak modularity.\footnote{You have reason to be confused here.  However, an orthomodular lattice is
in general not modular. Clearly a case of bad terminology, due to some formal confusion at the early development of the
subject, something what most probably did not contribute to its general appreciation.}  For ortholattices we have as
such the following hierarchy
${\rm Distributive}\ \Longrightarrow\ {\rm Modular}\ \Longrightarrow\ {\rm Weakly\ Modular}$\,,   
or, in terms of objects, 
${\sf BoolAlg}\subset{\sf MOL}\subset{\sf OML}$\,.  We refer to
Bruns and Harding (2000) for a recent survey on algebraic aspects of this matter. Finally recall that both examples
considered above are examples of so-called {\sl atomistic lattices} respectively having the one-dimensional subspaces
$\Sigma_{\cal H}$ and the singletons $\{\{x\}|x\in X\}$ as atoms, explicitly,
\beq
\forall A\in L_{\cal H}:A=\bigvee_{\cal H}\{ray(\psi)\in\Sigma_{\cal H}|\,ray(\psi)\subseteq A\} 
\quad\quad{\rm and}\quad\quad
\forall T\in {\rm P}(X): T=\bigcup\{\{x\}|x\in T\}\,, 
\eeq
thus satisfying the general {\sl atomisticity} condition $\forall a\in L:a=\bigvee\{p\in\Sigma|p\leq a\}$\,, where
$\Sigma$ denotes the atoms of $L$\,, i.e. $p\in\Sigma$ if and only if $\forall a\in L: a\leq p\Rightarrow a\in\{0,p\}$\,.
 
\smallskip  
Now, coming back to the two perspectives on logicality mentioned above, we will ``initially'' take the endo-perspective,
and look at the true `proper' qualities of the system, to which we will refer briefly as {\sl properties}. Later in the
paper, the exo-perspective will enter natually when defining logical hooks. The resulting structure will as such
incorporate both!  So in this paper a property is definitely not to be envisioned
merely as an observed quality/quantity, since that would be an event of the Foulis-Randall perspective. We as such do
assume a form of realism in the sense that properties do exist in absence of a measurement.\footnote{Reality is obviously
in no way to be understood as synonym for `locality and non-contextuality' as it is sometimes the case in some (from a
philosophical perspective) slightly naive discourses on philosophy of physics.} For example, in the dark, one could
attribute the property referred to as `red' to an object which is such that, ``whenever there is a white light source
brought in its environment that shines on it, it radiates red light''. Note here that we implicitly assume a system to be 
well-specified.  Depending on its possible realization  
$p$ (say {\sl state}), the system possesses different properties
$L_p$\,, referred to as the {\sl actual properties} for that particular realization $p$.  The collection of all
properties that the system can possess within the boundaries of its domain of specification, all the corresponding
realizations themselves being denoted as $\Sigma$ (any other realization will be considered as destruction of the
system), will be denoted by
$L$.  This set $L$ goes naturally equipped with a partial order in terms of ``actuality of $a\in L$ implies actuality
of $b\in L$'', i.e. for any (fixed) state we have that: if it is the case that ``whenever it (...) is in
environment 
$e_a$ then it causes phenomenon
$\alpha_a$ to happen'', then this implies that ``whenever it (...) is in environment $e_b$ then it causes phenomenon
$\alpha_b$ to happen''.  Denoting ``$a$ is actual in state $p$'' as $p\prec a$ this formally becomes
\beq\label{PropEquiv}
(a\leq b)\ \Longleftrightarrow\ (\forall p)(p\prec a\Rightarrow p\prec b)\,. 
\eeq  
Moreover, this poset is closed under
`conjunctions' $\bigwedge A$ of properties $A\subseteq L$ where actuality of $\bigwedge A$ stands for ``any $a\in A$ is
actual'', what actually means that, for each $a\in A$, whenever the system (...) is within environment $e_a$ then it
causes phenomenon  
$\alpha_a$ to happen.\footnote{Note here that contrary to a Tarskian perspective where one abstracts over the
true sense of meets, we give a particular operational significance to it.  Conjunctivity is in a sense
``conjunctivity with respect to actuality'', i.e. with respect to ``causing phenomena $\alpha_a$ for $a\in A$ to
happen whenever (...)''. See for example Girard, Lafont and Taylor (1989) for a survey of some similar operational
considerations on connectives in computation and proof theory, where one focuses in particular on the `dynamics'
underlying proofs and programs.}   So we consider here a not fully specified environment in order to establish, slightly
abusively, a disjunction of environments
$\{e_a\}_{a\in A}$ (and corresponding phenomena
$\{\alpha_a\}_{a\in A}$).\footnote{Or, in Piron's terms ``choose any $a\in A$ and place the system in $e_a$'', i.e. a
choice of environment.  Again, in order to avoid any cognitive connotation, we prefer to avoid the word `choice'
(although, we don't see any a priori problem in its use).} The feature that distinguishes quantum
systems from classical systems is the fact that we cannot define a disjunction of a collection
$A$ in this way.  Given $A\subseteq L$, the statement ``some  
$a\in A$ is actual'' would require a simultaneity, or again slightly abusively, a conjunction of environments what
conflicts with quantum theory where we have incompatibility of measurements corresponding to non-commuting
self-adjoint operators.  The conjunction $\bigwedge$ defined above provides $L$ with a complete lattice structure,
where, as already mentioned, the corresponding joins have not necessarily a disjunctive
significance.\footnote{`Disjunctive' to be seen again in terms of ``disjunctive with respect to actuality''.} In
particular can there be properties
$b\in L$ that do not imply that some $a\in A$ is actual but that do imply the join $\bigvee A\in L$ to be actual, the
so-called {\sl superposition principle} of quantum theory --- see Aerts (1981) and Coecke (2001a) for a rigorous
discussion on this matter.   Let us stress that at this point, as argued in Coecke, Moore and Smets (2001a), the full
physically derivable logical content that emerges from our operational setting consists of a consequence
relation\footnote{For the sake of the argument, we initially introduce here a Tarskian notion of consequence relation,
i.e.  following Tarski (1936, 1956)\,. As such it can be  seen as a binary relation on sets of formulas which satisfies 
reflexivity, monotonicity and transitivity.  Later on we will extend this notion of consequence relation to allow 
multiple conclusions --- see eq.(\ref{MultConc}) --- following  
the ideas of D. Scott. Note however that in contemporary literature this 
type of consequence relation is often 
replaced by a `weaker one' in the sense that substructural logicians   
prefer to work with multisets and/or non-monotonic 
logicians drop the monotonicity condition --- for more details on this 
matter we refer to Avron (1994).}  
$\vdash\
\subseteq {\rm P}(L)\times L$\,, that extends the lattice ordering $\leq\ \subseteq L\times L$ exploiting {\sl
conjunctivity}, i.e. \beq
\forall a\in A: ``a\ {\rm is\ actual}"\ \Longleftrightarrow\ ``\bigwedge A\ {\rm is\ actual}"\,,
\eeq   
since this allows to transcribe the set $\{``a\ {\rm is\ actual}"|\,a\in A\}$ as ``$\bigwedge A$ is
actual" it justifies setting\,\footnote{One could say that the notation $a, \ldots (a\in A)$ for representing
actuality of each member in $A$\,, i.e. $\forall a\in A: ``a\ {\rm is\ actual}"$ could be simplified by writing down
$A$\,, since in general in sequent calculus a list of assumptions on the left of $\vdash$ always has to be interpreted
conjunctively, i.e. as identifiable with the meet.  However, further we will consider collections $A\subseteq L$ in
terms of $\exists a\in A: ``a\ {\rm is\ actual}"$ and they will also appear on the left of $\vdash$ since we will
consider them as primitive propositions. The notion $a, \ldots (a\in A)$ is as such required to avoid confusion.}
\beq\label{ConjConseq}
a, \ldots (a\in A)\vdash b\ \Longleftrightarrow\ \bigwedge A\vdash b\ \Longleftrightarrow\ \bigwedge A\leq b\,.
\eeq
We can introduce a (semantic) satisfaction relation   
$\models\ \subseteq\Sigma\times L$\,, exactly being the actuality relation $\prec$ between states and properties --- we
prefer to have this double use of notation $\prec$ and 
$\models$ to stress whether we are either talking about the physical content or the derived logicality.  As
such satisfaction and consequence are for single assumptions related by  
\beq\label{Propconseq}
a\vdash b\ \Longleftrightarrow\ (\forall p)(p\models a\Rightarrow p\models b) 
\eeq
and thus in general we have 
that $a, \ldots (a\in A)\vdash b\ \Longleftrightarrow\ (\forall p)\bigl((\forall a\in A:p\models a)\Rightarrow
p\models b\bigr)$\,.  For transparancy of the argument
below we essentially consider single properties as arguments.  
The (hypothetical!) existence of some implication connective
$(-\to-): L\times L$ would at least require that it satisfies the so-called `minimal implicative condition',
for single assumptions being
$a\vdash b\ \Longleftrightarrow\ \vdash\!(a\to b)$\,,
such that it extends the physically derivable implication relation 
encoded as the lattice ordering, what transcribes in lattice and state terms respectively as   
\beq
a\leq b\ \Longleftrightarrow\ (a\to b)=1
\quad\quad{\rm and}\quad\quad
\forall p\in\Sigma:(p\prec a\Rightarrow p\prec b)\ \Longleftrightarrow\ \forall p\in\Sigma:p\prec (a\to
b)\,.
\eeq
%%% 
%% 
%
However, validity of {\sl deduction} moreover transcribes as ({\it sensu} Gentzen's sequent calculus) 
$\{a,c\}\vdash b\Rightarrow c\vdash\!(a\to b)$, or exploiting conjunctivity,
$a\wedge c\vdash b\Rightarrow c\vdash\!(a\to b)$\,,
what transcribes in lattice terms as
\beq\label{ForwHeytAdj}
a\wedge c\leq b\ \Longrightarrow\ c\leq\!(a\to b)\,. 
\eeq
Note here that the minimal implicative condition is actually a weakened form of the deduction theorem. 
Validity of both {\sl modus ponens}
$c\vdash\!(a\to b) \Rightarrow  \{c,a\}\vdash b$
assures the converse implication, i.e. \beq\label{BackwHeytAdj}
a\wedge c\leq b\ \Longleftarrow\ c\leq\!(a\to b)\,.
\eeq
% 
%%  
%%%
We will come back to this point in the next section after recalling adjointness.   

\bigskip\noindent
{\bf 3. SASAKI ADJUNCTION}     
     
\medskip\noindent
We recall some basic features of Galois adjoints. A more detailed survey of Galois adjoints can be found in Ern\'e et al
(1993) and for Galois adjoints in a more physical perspective we refer to Coecke and Moore (2000).
A pair of maps
$f^*:L\to M$ and
$f_*:M\to L$ between posets $L$ and $M$ is {\sl Galois adjoint},    
denoted by $f^*\dashv f_*$, if
and only if
\beq\label{eq:Gadj}
f^*(a)\leq b\Leftrightarrow a\leq f_*(b)\,.  
\eeq
Stressing the mathematical importance of
adjoints, we respectively quote the co-father of category theory S. Mac Lane and
logician R. Goldblatt (Goldblatt 1984 p.438):
\begin{quote}
``... adjoints occur almost everywhere in many branches of mathematics. ... a systematic use of all these
adjunctions illuminates and clarifies these subjects.'' \vspace{2mm} \\
``The isolation and explication of the notion of adjointness is perhaps the most profound contribution
that category theory has made to the history of general mathematical ideas.''
\end{quote}
One could even say that where in the beginning days of category theory the claim was made that it are the
functors and natural transformations that constitute the core of category theory rather than the categories
themselves, that it are actually the adjunctions that provide the true power.
Coming back to eq.(\ref{eq:Gadj}), in the case that $f^*$ and $f_*$ are inverse, and thus $L$ and $M$ isomorphic, the
above inequalities saturate in equalities. As we show below, the notion of Galois adjoint retains some
essential uniqueness properties of inverses. Whenever
$f^*\dashv f_*$ then $f^*$ preserves all existing joins and $f_*$ all existing meets.  This means that for a Galois
adjoint pair between complete lattices, one of these maps preserves all meets and the other preserves all
joins.  
Conversely, for $L$ and
$M$ complete lattices, any meet preserving map $f_*:M\to L$
has a unique join preserving left Galois adjoint
and any join preserving map $f^*:L\to M$ a unique meet preserving {right Galois adjoint}, respectively,
\beq\label{eqGalois2}
f^*:a\mapsto\bigwedge\{b\in M|a\leq f_*(b)\}\quad\quad\quad\quad f_*:b\mapsto\bigvee\{a\in L|f^*(a)\leq b\}\,.
\eeq
Thus, it follows that there is a one-to-one correspondence
between the join preserving maps between complete lattices and the meet preserving maps in the
opposite direction, this so-called `duality' being established by Galois adjunction.  One also verifies
that eq.(\ref{eq:Gadj}) is equivalent to  
\beq
\forall a\in L:a\leq f_*(f^*(a))\ \ {\rm and}\ \ \forall b\in M:f^*(f_*(b))\leq b\,.
\eeq 
Considering pointwise ordering of maps, i.e. for $f,g:L\to
M$, $f\leq g\Leftrightarrow \forall a\in L: f(a)\leq g(a)$ we can write the above as 
$id_L\leq f_*\circ f^*$ and $f^*\circ f_*\leq id_M$ 
where $id_L$ and $id_M$ are the respective identities on $L$ and $M$\,.  Now, coming back to
eq.(\ref{ForwHeytAdj}) and eq.(\ref{BackwHeytAdj}) of the previous section one sees that they define, when both of them
are valid, an adjunction $(a\wedge-)\dashv(a\to-)$ for all $a\in L$\,. This particular property, i.e. the existence
of a hook that acts as a right adjoint to the parameterized action of the meet, actually defines a Heyting algebra, a
type of lattice to which we turn our attention now. 

\smallskip 
Let us recall some features of Heyting algebras --- for a recent survey see for example
Borceux (1994). A {\sl Heyting algebra} is a lattice $(H,\wedge,\vee)$ equipped with an additional binary operation
${(-\to-)}: H\times H\to H$ that satisfies $a\wedge b\leq c\Leftrightarrow a\leq (b\to
c)$\,, i.e. after exchanging $a$ and $b$ and applying commutativity of $(-\wedge-)$\,, the action
of the meet is indeed left adjoint to the action of the hook, explicitly we have 
$(a\wedge-)\dashv(a\to-)$ for all $a\in H$\,. 
As such, a complete Heyting algebra encodes those lattices in which we encode validity of modus ponens and deduction in
a semantical way. Also following from this adjointness, in any Heyting algebra
$(a\wedge-)$ preserves existing joins, explicitly, $a\wedge(b\vee b')=(a\wedge b)\vee(a\wedge
b')$, so it turns out that a Heyting algebra is always distributive\,. In fact, any Boolean algebra, e.g.,
${\rm P}(X)$ for any set, turns out to be a Heyting algebra with $(a\to b)=\!^c a\vee b$\,, where
$^c$ denotes complementation, $(a\to b)$ then being logically interpretable as ``(not $a$) or $b$''. A
so-called {\sl pseudo-complement} can be defined on any Heyting algebra as
$\neg(-):H\to H:a\mapsto(a\to 0)$ given a lower bound
$0$ of
$H$\,. It then however turns out that contrary to a Boolean algebra we in general don't have that $\neg
a\vee a=1$ given an upper bound $1$ of $H$\,, what justifies the notion of pseudo-complement. 
In this paper we will only consider complete Heyting algebras, where a Heyting algebra is
complete if and only if the underlying lattice is complete. 
Now, since 
$(a\wedge-)$ preserves 
the joins of all subsets of a complete Heyting algebra $H$, we obtain a stronger form of distributivity namely
$a\wedge(\bigvee B)=\bigvee_{b\in B}(a\wedge b)$\,. In fact, this {\sl complete distributivity} now fully determines the
complete Heyting algebra structure in the sense that for all $a\in H$ the map
$(a\wedge-):H\to H$ preserves all joins so it has a unique right
adjoint ${(a\to-)}:H\to
H$. So complete Heyting algebras are complete lattices 
where the join of all
subsets is completely distributive over binary meets. 
We will now present an example of a complete Heyting algebra which is in general not a Boolean algebra. Let $L$ be
any poset and set   
$\downarrow\!a:=\{b\in L|b\leq a\}$  
for $a\in L$ and introduce a {\sl downset} or {\sl order ideal} as any set of the form
$\downarrow\![A]:=\{b\in L|\exists a\in A:b\leq a\}=\bigcup_{a\in A}\downarrow\!a$
with $\emptyset\not=A\subseteq
L$\,, i.e. $I$ is an order ideal 
if and only if $I \not = \emptyset$ and $a \leq b \in I$ implies $a \in I$\,. Order ideals of the form $\downarrow\! a$
for
$a\in L$ are called {\sl principal ideals}.  It then turns out that the collection of non-empty downsets 
${\rm I}(L):=\{\downarrow\![A]|\emptyset\not=A\subseteq L\}$
constitutes a complete Heyting algebra. Indeed, since unions and intersections of downsets are again
downsets, ${\rm I}(L)$ is closed under unions
and intersections from which it follows that they respectively constitute the join and meet in ${\rm I}(L)$\,. 
Distributivity of ${\rm I}(L)$ is as such inherited from that of ${\rm P}(L)$\,.   Using eq.(\ref{eqGalois2}) we can
now compute the corresponding Heyting algebra hook  
\beq
(B\to_{{\rm I}(L)}C)=\bigcup\{A\in {\rm I}(L)|A\cap B\subseteq C\}
=\{a\in L|\forall b\in B:a\wedge b\in C\}\,.
\eeq 
As pseudo-complement we obtain
$\neg B=(B\to_{{\rm I}(L)}0_{{\rm I}(L)})=\{a\in L|\forall b\in B:a\wedge b=0_L\}$\,.    
So in general we indeed do not have $B\bigcup\neg(B)=1_{{\rm I}(L)}$\,.
The most simple example of a Heyting algebra which is not Boolean is a
three element chain $\{0<a<1\}$\,. In particular are the downsets of any chain isomorphic to the chain itself,
establishing the claim that downsets in general don't constitute a Boolean algebra.  Another example are the open sets
of a topological space ordered by inclusion.    

\smallskip
Recall that for the lattice of closed subspaces of a Hilbert space the {\sl Sasaki projection}  
\beq\label{HilbertSasaki}
\varphi_A^*:L_{\cal H}\to L_{\cal H}:B\mapsto A\cap(A^\perp\vee_{\cal H} B)
\eeq
exactly encodes the action of the orthogonal projector $P_A$
that projects on the subspace $A$\,, so in particular we have for the action on rays that
\beq\label{HilbertSasakiBIS}
\varphi_A^*\bigl(ray(\psi)\bigr)=A\cap\bigl(A^\perp\vee_{\cal H}\, ray(\psi)\bigr)=ray\bigl(P_A(\psi)\bigr)
\eeq
where
we identify $ray\bigl(P_A(\psi)\bigr)$ in case that ${\psi\perp A}$ with the zero dimensional subspace. Moreover, for an
arbitrary orthomodular lattice $L$\,,  setting
\beq
\varphi_{a}^*:L\to L:b\mapsto a\wedge(a'\vee b)\quad\quad {\rm and}\quad\quad 
\varphi_{a,*}:L\to L:b\mapsto a'\vee(a\wedge b)\,,
\eeq
for all $a\in L$, we have $\varphi_a^*\dashv\varphi_{a,*}$\,.  Indeed, if $a\wedge(a'\vee b)\leq c$  
then $a'\vee\bigl(a\wedge
\left(a\wedge(a'\vee b)\bigr)\right)\leq a'\vee(a\wedge c)$ where 
$b\leq a'\vee b=a'\vee\bigl(a\wedge(a'\vee b)\bigr)$ since $a'\leq  
a'\vee b$\,, and analogously
one proves the converse. This adjunction actually embodies why the {\sl Sasaki hook}
$(-\stackrel{S}{\to}\,\cdot\,):=\varphi_{(-),*}(\,\cdot\,)$ has been interpreted as an implication, since  
$\varphi_a^*$ coincides with $(a\wedge -):L\to L$\,, the classical projections, in the case that $L$ 
is distributive since then $a\wedge(a'\vee b)=(a\wedge a')\vee(a\wedge b)=1\vee(a\wedge b)=a\wedge b$.\footnote{This view is obviously motivated by the
fact that where for a Heyting   algebra the actions
$\{(a\wedge-)|a\in L\}$ can be envisioned as projections on $a$,  
for orthomodular lattices the
Sasaki projections $\{\varphi_a^*|a\in L\}$ are the closed orthogonal projections in the Baer
$^*$-semigroup  of $L$-hemimorphisms (Foulis 1960) which, as mentioned above, coincide in the case of the subspace
lattice of a Hilbert space with the action of the closed projectors of the underlying Hilbert space. We also refer to
Coecke and Smets (2000) for complementary details on this matter.}  In particular do
we as such retain an adjunction of projection action and hook, mimicking the one that one has for complete Heyting
algebras that embodies the validity of modus ponens and the kind of deduction theorem obtained by combining
eq.(\ref{ForwHeytAdj}) and eq.(\ref{BackwHeytAdj})\,.  However, this in no way implies that all tools available in
classical/intuitionistic logic will still be valid within this setting. Let us briefly outline how the minimal implicative
condition and the adjointness for the Sasaki hook relate, both in the cases that we abstract over the explicit
formulation of the Sasaki hook and the Sasaki projection, i.e. the case of a general abstract adjoint implication
(Hardegree 1979, 1981) on a bounded poset, and the case of them being explicitly defined on an ortholattice.  We follow
Coecke, Moore and Smets (2001c). Let ${\rm J}(L)$ be the collection of isotone maps on a bounded poset $L$ that admit a
right adjoint.  An adjoint implication is then defined by a map
$\tilde{\varphi}^*:L\to {\rm J}(L):a\mapsto\tilde{\varphi}_a^*$ that satisfies $\tilde{\varphi}_a^*(1)=a$.  The
parameterized right adjoint  
$(-\stackrel{\tilde{\varphi}}{\to}-):L\times L\to L$\,, i.e. $\tilde{\varphi}^*_a\dashv (a\stackrel{\tilde{\varphi}}{\to}-):=\tilde{\varphi}_{a,*}$\,, is then to what we refer as
the {\sl adjoint implication}.  The condition $\tilde{\varphi}_a^*(1)=a$ implies the minimal implicative condition via
explicitation of the adjunction, i.e. $\tilde{\varphi}_a^*(1)\leq c\Leftrightarrow b\leq (a\stackrel{\tilde{\varphi}}{\to}c)$, for $b=1$. 
One could as such state that because the Sasaki projections satisfy
$\varphi_a^*(1)=a$\,, the Sasaki hook satisfies the minimal implicative condition.  For an ortholattice it
turns out that adjointness of the Sasaki hook and the Sasaki projection is equivalent to one side of the
implicative condition, namely $(a\stackrel{S}{\to}x)=1\Rightarrow a\leq x$ --- notice that the other side is trivially
satisfied for ortholattices since $a\leq x$ implies $a'\vee(a\wedge x)=a'\vee a=1$. From this
perspective one can say that the minimal implicative condition incarnates the fact that the Sasaki hook arises as the
right adjoint of Sasaki projections. Moreover, these two alternative definitions  are actually equivalent to
the ortholattice being orthomodular, and as such provide alternative characterizations of orthomodularity, respectively
one that can be written equationally, a rather logical one, and one in terms of an adjunction.  
\bpr
The following are equivalent for an ortholattice $L$\,:
\par\smallskip\noindent i. $L$ is orthomodular, i.e. $a\leq b$ implies $a\vee(a'\wedge b)=b$\,;
\par\smallskip\noindent ii. For all $a\in L$ we have $(a\stackrel{S}{\to}x)=1
\Rightarrow\,(or\, \Leftrightarrow)\,a\leq x$\,;
\par\smallskip\noindent iii. For all $a\in L$ we have $\varphi_a^*(-)\dashv(a\stackrel{S}{\to}-)$\,.  
\epr
%%%  
%% 
%
By (i.)\,$\Leftrightarrow$\,(ii.)\,, one has a statement concerning logicality attributed to the Sasaki hook in terms of
the minimal implicative condition, or equivalently, the Sasaki adjunction incarnates the utterance `orthomodular
logic'.\footnote{See also Moore (1993) on this matter.} As is reflected in the title and introduction of this paper, we do
not follow this line of thought! For us, it is (i.)\,$\Leftrightarrow$\,(iii.) that will provide a new interpretation of
orthomodularity in terms of causal duality.
% 
%%  
%%%

\bigskip\noindent
{\bf 4. TRUE IMPLICATIVE QUANTUM LOGICALITY}            
      
\medskip\noindent
In this section we essentially follow Coecke (2001a).  As discussed in section 2 the lattice of properties in general
does not encode arbitrary disjunction, since otherwise, they would constitute the joins and as such all joins would be
disjunctions, what is in general not the case.  Let us first analyze what happens in a (dichotomic) perfect quantum
measurement.\footnote{For the introduction of the respective concepts of (dichotomic) ideal measurement and
(dichotomic) measurement of the first kind, and, conjointly, a (dichotomic) perfect measurement, we refer to
Pauli(1958) and Piron (1976).}  Consider a (dichotomic) perfect quantum measurement of the property $a\in L$ and
correspondingly, its orthocomplement $a'$\,.  Assuming that a property
$b\in L$ is actual before the measurement, it follows, since the Sasaki projections encode projectors on subspaces,
that after the measurement either $\varphi_{a}^*(b)$ or $\varphi_{a'}^*(b)$ is actual.  Indeed, referring back
to eq.(\ref{HilbertSasaki}) and eq.(\ref{HilbertSasakiBIS})\,, and recalling that dichotomic measurements are in quantum
theory represented by self-adjoint operators with a  binary spectrum, the projectors on the corresponding (mutually
orthogonal) eigenspaces are then exactly encoded by $\varphi_{A}^*$ and $\varphi_{A^\perp}^*$\,, where $A$ and
$A^\perp$ are the corresponding eigenspaces.  Writing ``we obtain either $\varphi_{A}^*(ray(\psi))$ or
$\varphi_{A^\perp}^*(ray(\psi))$ as outcome state'' then corresponds to an abstraction over the corresponding
probabilistic weights of the two outcomes in a dichotomic measurement, focusing on the fact that whenever we
are not in an eigenstate, there is an uncertainty on the outcome. Consequently, there is also an uncertainty on the
corresponding `change of state' (according to the projection postulate), and as such, an uncertainty on the
corresponding `change of actual properties'.  We refer to this logical feature of quantum measurements as the
`emergence of disjunction in quantum measurements'. Writing this in a more formal, although intuitive way using a
consequence `symbol' we obtain:
\beq
``b\ {\rm actual}"\ \ \vdash_{{\rm perf.\ meas.\ of}\ \{a,a'\}}\ \ ``\varphi_{a}^*(b)\ {\rm actual}"\ \underline{\rm
or}\  ``\varphi_{a'}^*(b)\ {\rm actual}"
\eeq
or, when assuming the existence of an appropriate implicative connective $\longrightarrow_{{\rm perf.\ meas.\ of}\
\{a,a'\}}$ that satisfies the corresponding minimal implicative condition this becomes via the corresponding weakened
form of the deduction theorem:
\beq
``b\ {\rm actual}"\ \ \longrightarrow_{{\rm perf.\ meas.\ of}\ \{a,a'\}}\ \ ``\varphi_{a}^*(b)\ {\rm actual}"\
\underline{\rm or}\  ``\varphi_{a'}^*(b)\ {\rm actual}"\,.
\eeq
Unfortunately, $``\varphi_{a}^*(b)\ {\rm actual}"\ {\rm
or}\  ``\varphi_{a'}^*(b)\ {\rm actual}"$, i.e. ``a member of the pair $\{\varphi_{a}^*(b),\varphi_{a'}^*(b)\}$ is
actual'', is not encoded in the lattice of properties of a quantum system as an element since for example, taking $b=1$,
we have
$\varphi_{a}^*(1)\vee\varphi_{a'}^*(1)=1$ independent on
$a$ although the possible states the system can have --- given that  $``\varphi_{a}^*(b)\ {\rm actual}"\ {\rm
or}\  ``\varphi_{a'}^*(b)\ {\rm actual}"$ --- definitely
depend on $a$\,. Thus, it would make sense to have logical
propositions that express disjunctions of properties since they emerge in quantum processes, in the
endo-perspective. 
The question then arises whether we can extend 
$L$ with propositions of the type ``$\varphi_{a}^*(b)$ actual"
or $``\varphi_{a'}^*(b)$ actual", or equivalently, ``a member of $\{\varphi_{a}^*(b),\varphi_{a'}^*(b)\}$ is
actual'', without loosing the logicality encoded in the initial
lattice of properties, i.e. the lattice order, and whether this can be done in a non-redundant, canonical or even
mathematically universal way.\footnote{Note that the fact that the \underline{or} that we obtain in a perfect
measurement is exclusive does not have to be encoded explicitly since it is already captured by the
orthocomplementation since we have
$\varphi_{a}^*(b)\wedge\varphi_{a'}^*(b)=0$ by $a\wedge a'=0$ such that ``both $\varphi_{b}^*$ and $\varphi_{b'}^*$ are
actual'' is excluded --- $0$ indeed encodes the `absurd'.}    

\smallskip
A first candidate for encoding disjunctions would be the powerset ${\rm P}(L)$. However, if
$a\leq b$ we don't have $\{a\}\subseteq\{b\}$ so we do not preserve the initial logicality, or, otherwise stated, if
$a<b$ then the {\sl propositions}
$\{a\}$ and $\{a,b\}$ (`read' $\{a,b\}$ as: either $a$ or $b$ is actual) mean the same thing, since actuality
of $b$ is implied by that of $a$\,. We can clearly overcome this problem by restricting to order ideals
${\rm I}({L}):=\{\downarrow\![A]|A\subseteq{L}\}\subset {\rm P}(L)$\,. However, we encounter a second problem.  In
case the property lattice would be a complete Heyting algebra in which
all joins encode disjunctions, then $A$ and $\{\bigvee A\}$ again mean
the same thing. As argued in Coecke (2001a), this redundancy is then exactly eliminated by considering
distributive ideals ${\rm DI}(L)$ (Bruns and Lakser 1970), that is, order ideals, that are closed under {\sl joins of
distributive sets} (abbreviated as {\sl distributive
joins}), i.e. if $A\subseteq I \in {\rm DI}(L)$ then $\bigvee A \in I$ whenever  we have $\forall b \in {L}: b 
\wedge\bigvee A = \bigvee\{b \wedge a \mid a \in A \}$.  For ${L}$ atomistic and
$\Sigma \subseteq {L}$\,, ${\rm DI}(L) \cong {\rm P}(\Sigma)$ which
implies that ${\rm DI}(L)$ is a complete atomistic Boolean algebra  
(Coecke 2001a)\,. 
We can moreover provide a from a mathematical perspective more regourous reasoning which exhibits the canonical nature
of this construction.  Consider the following definitions for $A\subseteq L$\,:   
\par\smallskip\noindent
i. $\bigvee A$ is called {\sl disjunctive} iff (``$\bigvee A$ is actual" $\Leftrightarrow$ $\exists a\in A: ``a$
is actual")\,;\footnote{Compare this definition with the one of conjunctivity.}  
\par\smallskip\noindent
ii. {\sl Superposition states} for $\bigvee A$ are states for which ``$\bigvee A$ is actual" while ``no $a\in A$ is
actual"\,;   
\par\smallskip\noindent
iii. {\sl Superposition properties} for $\bigvee A$ are properties of which the actuality implies that ``$\bigvee A$
is actual'' (without being equivalent to $\bigvee A$)\,, and, that can be actual while ``no $a\in A$ is actual".
\par\smallskip\noindent
%%% 
%% 
%
Extending the satisfaction relation encoding actuality by $p\models A\ \Longleftrightarrow\ \exists a\in A:p\models
a$ allows us  to set
\beqn 
(\forall p)\bigl(p\models \bigvee A\, \Leftrightarrow\,
p\models A\bigr)&\Longleftrightarrow&\bigvee A\ {\rm is}\ {\rm disjunctive}\,,\\
(p\models \bigvee A\ \ and\ \ p\not\models A)&\Longleftrightarrow&p\in\Sigma\ {\rm is\ a\ superposition\ state\ of}\
\bigvee A\,,\\ 
(b\,\dot{\vdash}\,\bigvee
A\ \ and\ \ b\not\vdash A)&\Longleftrightarrow&b\in L\ {\rm is\ a\ superposition\ property\ of}\ \bigvee A\,.  
\eeqn  
where $a\,\dot{\vdash}\,b$ means $a\vdash b$ but $b\not\vdash a$\,.
% 
%%  
%%%
\bpr
If ``existence of superposition states implies existence of superposition properties" then 
\beq
\bigvee A\ disjunctive\
\Longleftrightarrow\ \bigvee A\ distributive\,. 
\eeq
\epr
\bpf
See  Coecke (2001a)\,.
\epf
The necessity condition ``existence of superposition 
states implies existence of superposition properties'' can be illustrated by means of the 
following example.  When considering the four element lattice ${\{0\leq a,a'\leq 1\}}$ even then distributivity and
disjunctivity are not necessarily equivalent, for example in the case that $\Sigma=\{p,q,r\}$ and $p\prec a$,
$q\prec a'$,
$r\prec 1$ and $r\not\prec a,a'$ --- the superposition state $r$ for $a\vee a'$ has no corresponding superposition
property. Note here that for the particular example of quantum theory the condition is trivially
satisfied as it is the case for any atomistic property lattice.\footnote{Whenever this condition is satisfied one can
also construct a disjunctive extension, which is obviously not anymore the distributive extension, but something
`inbetween' the distributive extension and the downset completion where corresponding inclusions are injective order
embeddings that preserve all meets and the bottom element (Coecke 2001a).} Now, any complete lattice
$L$ has ${\rm DI}(L)$ of distributive ideals as its distributive hull (Bruns
and Lakser 1970), providing the construction with a (quasi-)universal property.\footnote{Injective hulls are
actually not universal in a strictly categorical sense.  However, it is possible to give a characterization of
distributive hulls in terms of a so-called `frame completion', which is a monoreflection and as such strictly
universal. See for example Harding (1999) and Stubbe (2001).} 
Moreover, ${\rm DI}(L)$ itself always proves to be a complete Heyting algebra and  
the inclusion preserves all meets and existing 
distributive joins. Thus,
${\rm DI}({L})$ encodes all possible disjunctions of properties, and moreover, it turns out that all
${\rm DI}({L})$-meets are conjunctive and all
${\rm DI}({L})$-joins are disjunctive --- note again that this is definitely not the case in the powerset ${\rm P}(L)$
of a property lattice, since $\{a\}\cap\{b\}=\emptyset$ whenever $a\not=b$ independent of what $a\wedge b$ is.
This means that we can `extend' the consequence relation of eq.(\ref{ConjConseq}) to $\vdash\
\subseteq{\rm P}({\rm DI}(L))\times{\rm P}({\rm DI}(L))$ respectively in terms of consequence and satisfaction as
\beqn\label{MultConc}
A, \ldots (A\in {\cal A})\vdash B, \ldots (B\in {\cal B})&\Longleftrightarrow&\bigcap{\cal A}\vdash\bigvee_{{\rm
DI}(L)}{\cal B}
\ \Longleftrightarrow\ 
\bigcap{\cal A}\subseteq\bigvee_{{\rm DI}(L)}{\cal B}\,,\\
& \Longleftrightarrow& (\forall p)\bigl((\forall A\in
{\cal A}:p\models A)\Rightarrow (\exists B\in {\cal B}:p\models B)\bigr)
\eeqn 
where we recall that $a\leq b\,\Leftrightarrow\,\downarrow\!a\subseteq\downarrow\!b$ and as such
$\downarrow\!\left(\bigwedge A\right)=\bigcap_{a\in A}\!\downarrow\!a$ encodes the properties within this set
of propositions on properties --- $A\subseteq L$ is here to be seen as just a set of properies without the disjunctive
connotation of the distributive ideals in ${\rm DI}(L)$\,. As
demonstrated in Coecke (2001a) it follows from all this that the object equivalence between:
\par\smallskip\noindent
i. complete lattices, and, 
\par\smallskip\noindent
ii. complete Heyting algebras
equipped with a {\sl distributive closure}, i.e. it preserves distributive sets,  
\par\smallskip\noindent
encodes a disjunctive representation for property lattices, where we, ${\rm DI}(L)$ being a complete Heyting algebra, do
have a Heyting hook 
\beq
(-\to_{{\rm DI}(L)}-):{\rm DI}(L)\times{\rm DI}(L)\to{\rm DI}(L):(B,C)\mapsto
\bigvee_{{\rm DI}(L)}\{A\in{\rm DI}(L)|A\cap B\subseteq C\}\,,
\eeq
where the explicit ${\rm DI}(L)$-joins are given by
\beq
\bigvee_{{\rm DI}(L)}:{\rm P}({\rm DI}(L))\to {\rm DI}(L):{\cal A}\mapsto  
\bigcap\left\{B\in{\rm DI}(L)\Bigm|B\supseteq\!\bigcup{\cal A}\right\} 
\eeq
(meets are obviously intersections), and we will argue below that this hook is {\sl implicative} sensu ``with
respect to actuality'', i.e. in the sense that in the property lattice $\bigwedge$ was conjunctive and in the sense we
defined disjunctivity when motivating the use of distributive ideals. 
One verifies that we actually obtain\footnote{Note here that
$\{a\in L\,|\,\forall b\in B:a\wedge b\in C\}$ indeed defines a member of ${\rm DI}(L)$\,.  First, we have that
$x'\leq x$ implies $x'\wedge b\leq x\wedge b$ for $x\in (B\to_{{\rm DI}(L)}C)=\{a\in L\,|\,\forall b\in B:a\wedge b\in
C\}$ such that 
$x'\wedge b\in C$ since $C\in{\rm DI}(L)$\,. Next, for $X\subseteq(B\to_{{\rm DI}(L)}C)$ with $X$ a
distributive set, i.e. $\forall c\in L: c\wedge\bigvee X= \bigvee\{c \wedge x \mid x \in X \}$\,, one easily verifies
that distributivity of $X$ implies distributivity of $\{b\wedge x \mid x \in X \}$ and thus $\left(\bigvee
X\right)\wedge b=\bigvee\{b \wedge x \mid x \in X \}\in C$ so $\bigvee X\in(B\to_{{\rm DI}(L)}C)$\,.}
\beq\label{ImpEquProp}
(B\to_{{\rm DI}(L)}C)=\{a\in L\,|\,\forall b\in
B:a\wedge b\in C\}\,.
\eeq  
The true quantum features are (at this `static' level) encoded in an {\sl operational resolution}
\beq
{\cal R}_{{\rm DI}(L)}:{\rm DI}(L)\to{\rm DI}(L):A\mapsto\bigm{\downarrow}\!\Bigl(\bigvee_L A\Bigr)
\eeq
that recaptures statements expressing actuality of properties within the larger collection of propositions on
actuality of them, and that only for classical systems becomes trivial, being the identity. The logical essence of this
representation is such that, rather than seeing the shift ``from classical to quantum'' as a weakening of the property
lattice structure from a distributive lattice to a non-distributive one, we envision this transition as going from a
trivial additional operation on the propositions (which as a consequence in the classical case coincide with the
properties) to a non-trivial one.  Note that the non-distributive features are as such recaptured as the range of this
additional operation ${\cal R}$\,, but they don't affect distributivity of the domain.  Quantum logic becomes as such
ordinary logic with an additional operation, a bit in the sense of modal logic. 
From a
pragmatic formal attitude, this construction however  seems to conflict with statements about the non-distributive
nature of quantum theory, what, for some authors is exactly the essence of quantum logicality.  In the quantum case, the
non-distributivity does not come in within the ordering of propositions, but as the range of the operation
${\cal R}$ which acts on the propositions. Also the other axioms considered in axiomatic approaches, e.g.
orthomodularity, have the same incarnation. We refer to Coecke (2001b) for a more elaborated discussion on the
significance/conceptions of non-distributivity in the context of quantum theory.

\smallskip
Going back to the explicit construction of the Heyting hook for propositions on properties, it as such 
also turns out that the canonical implication on a lattice of properties is an external one that takes values in
${\rm DI}(L)$, namely  
\beq\label{ImpForProp}
(-\to_L-) :\ L\times L\to {\rm DI}(L):(b,c)\mapsto\{a\in L|a\wedge b\leq c\} 
\eeq
obtained by domain and codomain restriction of $(-\to_{{\rm DI}(L)}-)$\,. If and only if
$L$ is itself a complete
Heyting algebra, then we can represent this external operation
faithfully as an internal one by
setting $(b\to c):=\bigvee(b\to_L c)$\,. In
particular can our external
implication arrow be defined by
\beq\label{PseudoAdj}
a\wedge b\leq c\ \Longleftrightarrow\ a\in(b\to_L c)\,,  
\eeq
as such in a more explicit manner expressing that it generalizes the
implication that lives on
a complete Heyting algebra where $a\in(b\to_L c)$ then coincides with 
$a\leq(b\to c)$\,. 
Within ${\rm DI}(L)$ we see that $(b\to_L c)$ is the set of properties whose actuality makes the deduction ``if $b$
is actual then $c$ is actual'' true, i.e. given $a\in(b\to_L c)$\,, then $\forall p\in\Sigma_a: (p\models b\Rightarrow
p\models c)$ where $\Sigma_a:=\{p\in\Sigma|p\models a\}$\,.   In other words, $(-\to_L -)$ transcribes in terms of
actuality the minimal requirement of any functional formal implication with respect to extensional quantification over
the state set.   Note that by constructing $(-\to_L-)$ via (slightly abusively) domain restriction from ${\rm
DI}(L)\times{\rm DI}(L)$ to $L\times L$ we exhibit clearly that $(-\to_L-)$ can (again) be made
internal via a domain extension, and, that this extension has physical significance and moreover prerserves all the
physically derivable logicality of $L$\,.  As a statement:  ``$(-\to_L-)$ is the [closest you can get to] implication
on the lattice of properties''. In relation to the minimal implicative condition we obtain 
\beq\label{ImplExt1}
{L} \in (a \to_{L} b)\ \Longleftrightarrow\ a \leq b\ \Longleftrightarrow\ \forall p \in \Sigma: (p \models a
\Rightarrow p\models b)
\eeq
for $(-\to_L-)$ whereas for $(-\to_{DI({L})}-)$ this `extends' to  
\beq\label{ImplExt2}
L=(A \to_{{\rm DI}({L})}B)\ \Longleftrightarrow\ A \subseteq B\ \Longleftrightarrow\ \forall p \in \Sigma: (p \models A
\Rightarrow p\models B)\,.
\eeq
Note that $\to_{{\rm DI}({L})}$ as an
operation, is the parameterized right adjoint with respect to the respective meet actions ${\{(A\cap-)|A\in {\rm
DI}({L})\}}$\,. The above leads to a semantical interpretation of $(-\to_{{\rm DI}({L})}-)$ as 
${p \models (A \to_{{\rm DI}({L})} B)}\ \Longleftrightarrow\ (p \models A \Rightarrow p \models B)$\,,    
or, equivalently,  
\beq\label{eq:ImpVia Kripke}
\mu(A\to_{{\rm DI}({L})} B)=\{p\in\Sigma|p \models A \Rightarrow p \models B\}\,, 
\eeq
where, extending the usual {\sl Cartan map} $\mu:L\to{\rm P}(\Sigma):a\mapsto\{p\in\Sigma|p\prec a\}$\,,
we define
$\mu(A):=\bigcup\mu[A]$\,, the square brackets referring to pointwise application of $\mu$\,, i.e. in semantical terms,
$\mu(A)={\{p\in\Sigma|p\models A\}}$\,. Note at this point that there is indeed a duality in representing propositions
in terms of ${\rm DI}(L)$ or in terms of a particular subset ${\rm F}(\Sigma)$ of ${\rm P}(\Sigma)$ defined as ${\rm
F}(\Sigma):=\{\mu(A)|A\in{\rm DI}(L)\}$\,, which, for atomistic
$L$\,, turns out to be ${\rm P}(\Sigma)$ itself. In more syntactical terms, i.e. without referring to the state space
interpretation, adjointness allows us explicitly to restate 
$(-\to_{{\rm DI}({L})}-)$ as\,\footnote{Since $A\subseteq B\Longleftrightarrow \forall a\in A:a\in
B\Longleftrightarrow \forall p\models A:p\models B \Longleftrightarrow(\forall
p)(p\models A\Rightarrow p\models B)$\,, due to the equivalence induced by eq.(\ref{Propconseq})\,, we have
$C\subseteq(A\to_{{\rm DI}({L})} B)\Longleftrightarrow (C\cap A)\subseteq B
\Longleftrightarrow(\forall p)(p\models C\cap A\Rightarrow p\models B)
\Longleftrightarrow(\forall p)(p\models C, p\models A\Rightarrow p\models B) 
\Longleftrightarrow\forall p\models C:(p\models A\Rightarrow p\models B)$\,, 
where again eq.(\ref{Propconseq}) allows expressing this in terms of
$D\vdash C$ rather than $p\models C$\,.} 
\beqn 
(A\to_{{\rm DI}({L})} B) \!\!&=&\!\! \bigvee_{{\rm DI}({L})} \{ C \in {\rm DI}({L}) \mid \forall D\vdash C:(D \vdash A
\Rightarrow D
\vdash B)\}\\
\!\!&=&\!\! \bigvee_{{\rm DI}({L})} \{ C \in {\rm DI}({L}) \mid \forall d\in C:(d\in A \Rightarrow d\in
B)\}\\
\!\!&=&\!\! \{ c\in L \mid \forall d\leq c:(d\in A \Rightarrow d\in  
B)\}\,, 
\eeqn
for the latter, rather than explicitly showing that is indeed a distributive ideal such that we can drop the
corresponding closure, we can use eq.(\ref{ImpEquProp}) and verify straightforwardly that 
\beq
\forall a\in A:a\wedge c\in B\ \Longleftrightarrow\ \forall d\leq c:(d\in A\Rightarrow d\in B)\,. 
\eeq
As such, for properties we obtain, syntactically, 
\beq
(a\to_L b)=\{ c\in L \mid \forall d\vdash c:(d\vdash a \Rightarrow d\vdash b)\}\,.  
\eeq 
How does the
Sasaki hook relate to this implication? 
Since $\varphi_a^*(b)=a\wedge(b\vee a')\geq a\wedge b$ and thus $\varphi_-^*(-)\geq (-\wedge-)$ pointwisely --- recall
here that $(-\stackrel{L}{\to}-)$ arises as domain restriction of the right adjoint of the action of the ${\rm
DI}(L)$-meets, the latter encoding for properties the $L$-meet in terms of intersection of principal ideals. For the
corresponding adjoints of the actions we have
$a\wedge(a\stackrel{S}{\to}b)=a\wedge(a'\vee(b\wedge a))=
\varphi_a^*(b\wedge a)=b\wedge a\leq b$\, and thus $(a\stackrel{S}{\to}b)\in\{c\in L|a\wedge c\leq
b\}=(a{\to}_Lb)$\,, or, differently put, $\downarrow\!(-\stackrel{S}{\to}-)\leq(-{\to}_L-)$
pointwisely. In terms of actions this gives us for the corresponding adjoint pairs $(a\wedge-)\dashv(a{\to}_L-)$
--- with slight abuse of notation, see eq.(\ref{PseudoAdj}) --- and
$\varphi_a^*(-)\dashv(a\stackrel{S}{\to}-)$ 
\beq
(a{\to}_L-)\,\geq\,\downarrow\!(a\stackrel{S}{\to}-)\quad\quad\quad\quad{\rm and}\quad\quad\quad\quad
(a\wedge-)\,\leq\,\varphi_a^*(-)\,,
\eeq
expressing reversal of pointwise order by adjunction.  Thus, from a semantical perspective,
$\downarrow\!(-\stackrel{S}{\to}-):L\times L\to{\rm DI}(L)$ is a restriction of the (static) implication $(-\to_L-)$\,.
So, if the Sasaki adjunction doesn't encode a `real' implication, what does it do.  This will be
explained in the next section. 

\bigskip\noindent
{\bf 5. THE SASAKI ADJUNCTION INCARNATES CAUSAL DUALITY}       

\medskip\noindent 
Causal duality has been derived in Coecke, Moore and Stubbe (2001) inspired on derivations in Faure, Moore and Piron
(1995).  Rather than giving a full derivation, we sketch a more intuitive way of looking at the obtained results.  
Assume (so we don't give a full proof here) for a system placed in an environment $e$\,, e.g., an environment $e_a$
sensu section 2,\footnote{And for simplicity assumed to be non-destructive.} during a time interval $[t_1,t_2]$ (which
can be envisioned as being infinitesimal) that there exist the maps:\footnote{The existence can be proved --- see
Coecke,  Moore and Stubbe (2001) and Faure, Moore and Piron (1995).}
\par\smallskip\noindent
i. `Propagation of properties' $e^*:L_1\to L_2$ that assigns to any property $a_1\in L_1$ the strongest property
$e^*(a_1)\in L_2$ of which actuality is implied at time $t_2$ due to actuality of $a_1$ at time $t_1$\,;
\par\smallskip\noindent
ii. `Causal assignment of properties' $e_*:L_2\to L_1$ that assigns to any property $a_2\in L_2$ the weakest property
$e_*(a_2)\in L_1$ whose actuality at time $t_1$ guarantees actuality of $a_2$ at time $t_2$\,.
\par\smallskip\noindent
Since, given $a_2\in L_2$\,, $e_*(a_2)\in L_1$ guarantees actuality of $a_2$ at time $t_2$\,, $e_*(a_2)$ has to 
propagate to a property that is stronger (or equal) than $a_2$ and as such $e^*(e_*(a_2))\leq a_2$\,. Analogously, given
$a_1\in L_1$\,, since it propagates into $e^*(a_1)$\,, actuality of $a_1$ at $t_1$ guarantees actuality of $e^*(a_1)$ at
$t_2$ and as such $a_1\leq e_*(e^*(a_1))$\,.  Thus, from $e^*(e_*(a_2))\leq a_2$ and $a_1\leq e_*(e^*(a_1))$ we obtain
$e^*\dashv e_*$\,, and this adjunction is what we refer to as causal duality. The generality of the principle lies in
the fact that besides applying to temporal processes it also applies to compoundness (Coecke 2000).\footnote{We want
to stress here that causal duality actuality allows us to prove things and is such is not just a fancy way of
writing things down. For a proof of linearity of Schr\"odinger flows, given that the property lattice of the
corresponding system is $L_{\cal H}$\,, see Faure, Piron and Moore (1995)\,. For a proof that the tensor product of
Hilbert spaces is appropriate to describe compoundness for systems with as property lattice $L_{\cal H}$ see
Coecke (2000)\,. Conclusively, if the space in which we describe the system is linear, then causal duality
forces temporal propagation and compoundness to be described by linear maps. These results essentially use
Faure and Fr\"olicher (1993, 1994)\,.}     
 
\smallskip 
We started the first paragraph of the previous section with a discussion on (dichotomic) perfect quantum
measurements with the aim to exhibit the emergence of disjunction.  In view of section 2 we can denote the
corresponding environment that provokes such a measurement, i.e. the presence of the corresponding measuring device, as
$\varphi_{\{a,a'\}}$\,.  In the following paragraphs we then additionally argued that the
disjunctive extension has the extra advantage that it allows us to encode an external implicative hook on $L\times L$
which then extends to an internal implication on the whole of ${\rm DI}({L})\times{\rm DI}({L})$\,.  Thus, the use of
the disjunctive extension for representing quantum systems goes beyond representing the emergent disjunction in the
sense that it has also a pure logical motivation in terms of envisioning (static) quantum logicality as ordinary
logicality with the additional presence of a non-trivial operational resolution ${\cal R}:{\rm
DI}({L})\to{\rm DI}({L})$\,.  For the particular case of a perfect quantum measurement, we are going to restrict us now
to the specific example where we consider a transition only provided a certain positive outcome is
obtained, say $a$ for simplicity, what actually means that whenever the system is within environment
$\varphi_{\{a,a'\}}$ we condition on the fact that $a$ is obtained --- note however, not
necessarily in a causal manner, i.e. $a$ does not have to be actual before the measurement. A concrete way to envision
this specific situation is in terms of a filter, that whenever the outcome corresponding to $a$ is not obtained the
system will be destroyed.\footnote{See also Piron (1976) on measurements as filters and see Smets (2001) for a recent
survey.}  Let us denote the corresponding environment as
$\varphi_a$\,. Now, since the Sasaki projection $\varphi_a^*:L_{(1)}\to
L_{(2)}$ --- $t_2=t_1+\epsilon$ --- encodes the behavior of a system under a perfect quantum measurement
$\varphi_{\{a,a'\}}$ when the outcome corresponding to $a$ is obtained, it encodes the propagation of properties
with respect to $\varphi_a$ (justifying the notation $\varphi_a^*$ in perspective of the previous paragraph) and should
as such admit a left adjoint expressing (backward) causal assignment, and this is exactly how
$(a\stackrel{S}{\to}-)=\varphi_{a,*}(-):L_{(2)}\to L_{(1)}$ arises in this setting. Sasaki adjunction constitutes as
such an incarnation of causal duality.  This already `partly' explains the title of this paper --- a more compelling
perspective will be discussed in the next section. In particular will we show how causal duality extends to a dynamic
logical setting.  First we need to introduce causal relations as a dynamic counterpart to the static ordering of
properties in the property lattice.   
  
\smallskip    
Along the lines of the heuristics behind eq.(\ref{PropEquiv}) we can introduce the following two relations, whenever
an environment $e$ is specified, {\sl taking a $t_1$-perspective}\,:
\beqn
&&\stackrel{e}{\leadsto}\ \subseteq L_1\times L_2: a_1\stackrel{e}{\leadsto}a_2\,\Leftrightarrow\ {\rm``actuality\
of}\ a_1\ {\rm at}\ t_1\ {\rm implies}\ \Box{\rm \mbox{-}actuality\ of}\ a_2\ {\rm at}\ t_2\ {\rm''}\,;\\
&&\stackrel{e}{\looparrowleft}\ \subseteq L_1\times L_2: a_1\stackrel{e}{\looparrowleft} a_2\,\Leftrightarrow\
``\Box{\rm \mbox{-}actuality\ of}\ a_2\ {\rm at}\ t_2\ {\rm implies\ actuality\ of}\ a_1\ {\rm at}\ t_1{\rm ''}\,.
\eeqn
Now, what do we mean by {\sl taking a $t_1$-perspective}\,, and, {\sl $\Box$-actuality}?  From
the perspective at time $t_1$\,, i.e. before the interaction of the system and the environment $e$ takes place, there
are two modes of envisioning actuality at time $t_2$\,, namely i. ``$a_2$ can be actual'',
the uncertainty being due to the indeterministic nature of the interaction of the system with the environment, and,
ii. ``$a_2$ will be actual'', definitely.  Note that for
deterministic transitions these two coincide.  Motivated by the modal logic symbolism, we can refer to these two
alternatives respectively as {\sl  $\Diamond$-actuality} and {\sl
$\Box$-actuality}\,\footnote{For the use of 
modal-operators in static operational quantum logic, 
where the operators point out the so-called `classical limit properties' we refer to 
Coecke, Moore and Smets (2001a) and Smets 
(2001, \S 10)\,. This however shouldn't be confused with the association to modalities made in this paper.} ---
whenever we mention {\sl
$\Diamond$-actuality} and {\sl
$\Box$-actuality} we implicitly refer to a $t_1$-perspective.\footnote{Note here that the notions {\sl
$\Diamond$-actuality} and {\sl
$\Box$-actuality} have only significance with respect to a $t_1$-perspective. In particular, referring to the two modes
of envisioning actuality at time $t_2$ in the $t_1$-perspective, in a
$t_2$-perspective there is only one since the interaction of the system with the environment did take place.} In general
we clearly have for a fixed property $a_2$ at 
$t_2$ that
$\Box{\rm -actuality}\ \Longrightarrow\ \Diamond{\rm -actuality}$\,.  
Formally, this gives
us the following (semantical) definitions for $\stackrel{e}{\leadsto}$ and
$\stackrel{e}{\looparrowleft}$\,:
\beqn\label{semdefcaqusrel1}
a_1\stackrel{e}{\leadsto}a_2\ \Longleftrightarrow\ (\forall p)_1(p\models a_1 \Rightarrow \tilde{e}^*(\{p\})\models
a_2)\\
\label{semdefcaqusrel2}
a_1\stackrel{e}{\looparrowleft} a_2 \ \Longleftrightarrow\ (\forall p)_1(p\models a_1 \Leftarrow
\tilde{e}^*(\{p\})\models a_2)
\eeqn
where the index $1$ in $(\forall p)_1$ refers to the fact that we quantify over states at time $t_1$\,, where
$\tilde{e}^*(\{p\})$ denotes the states the system can have after interaction with the environment $e$ and where
$\tilde{e}^*(\{p\})\models
a_2$ stands for $\forall q\in\tilde{e}^*(\{p\}):q\models a_2$\,.  More explicitly referring to eq.(\ref{PropEquiv})
we see that as such the  relations
$\stackrel{e}{\leadsto}$ and
$\stackrel{e}{\looparrowleft}$ can be defined in terms of the actuality relation. The major advantage of taking
an {\sl a priori} $t_1$-perspective is that it will allow us to introduce binary connectives that extend this
relation `with the same codomain', this extension is to be envisioned in the sense that the relation
$\leq\ \subseteq L\times L$ has been extended to an implication --- {\it sensu} eq.(\ref{ImplExt1}) and eq.(\ref{ImplExt2})
--- namely
$(-\to_{{\rm DI}(L)}-)$\,, provided that we considered the disjunctive extension ${\rm DI}(L)$ of $L$ and not just 
$L$ itself.  When asking the question whether in some manner the relations $\stackrel{e}{\leadsto}$ and 
$\stackrel{e}{\looparrowleft}$ indeed extend to connectives it will as such be no surprise that we should again rather
consider ${\rm DI}(L_i)$ than $L_i$ itself.  By `with the same codomain' we mean that both will be represented in 
${\rm DI}(L_i)$ --- note here indeed that we do not require $L_1\cong L_2$ and as such also not ${\rm
DI}(L_1)\cong{\rm DI}(L_2)$\,.  
%%%%%
%%%%
%%% 
%%  
%
(Obviously, all this requires to some extend a pluralistic attitude, we indeed admit that: One could for example find
a motivation to consider a $t_2$-perspective; this then leads us to a bouquet of definable
causal relations and corresponding dynamic implications.)  
% 
%%  
%%%
%%%%
%%%%%
Note also that the necessity of
having to consider the disjunctive extension already follows from the following observation: given a specified
referential frame, one could define an environment {\it freeze} (with obvious significance), for which it clearly
follows that   
$(-\stackrel{{\it freeze}}{\leadsto}-)\ \equiv\ \ (-\leq-)\ \ \equiv\ (-\stackrel{{\it freeze}}{\looparrowleft}-)$\,,
i.e. {\it freeze} provides a {\sl static limit} for the more general {\sl dynamic formalism} that involves
explicitation of the environment.  As it can be seen in eq.(\ref{semdefcaqusrel1}) and eq.(\ref{semdefcaqusrel2})\,,
all these considerations will involve introducing the notions of propagation of propositions and causal
assignment for propositions, or equivalently, in terms of the corresponding sets of states that make a proposition true
with respect to actuality of one of its members, e.g., $\tilde{e}(\{p\})$\,.  We will do this in the next paragraph. 
First we take a look on how these relations are realized for the above discussed heuristics for the Sasaki
adjunction.
Following Coecke, Moore and Stubbe (2001)\,, we obtain respectively by the definitions of
$\varphi_a^*$ and
$\stackrel{\varphi_a}{\leadsto}$\,, and, explicit expression of causal duality, that\,\footnote{Indeed,
$\varphi_a^*(a_1)\leq a_2\
\Longrightarrow\ a_1\stackrel{\varphi_a}{\leadsto}a_2$ follows from the definition of $\leq$\,, and
$\varphi_a^*(a_1)\leq a_2\ \Longleftarrow\ a_1\stackrel{\varphi_a}{\leadsto}a_2$ follows from the fact that
$\varphi_a^*(a_1)$ is the strongest property who's actuality is implied by that of $a_1$ and as such implies any other
of that kind.}
\beq\label{SasakiLeadsto}
\varphi_a^*(a_1)\leq a_2\ \Longleftrightarrow\ a_1\stackrel{\varphi_a}{\leadsto}a_2\ \Longleftrightarrow\ 
a_1\leq \varphi_{a,*}(a_2)\,,
\eeq 
from which it also follows that $a_1\stackrel{\varphi_a}{\leadsto}\varphi_a^*(a_1)$\,, that
$\varphi_{a,*}(a_2)\stackrel{\varphi_a}{\leadsto}a_2$\,, and in particular, using $\varphi_a^*(1)=a$\,, that
$1\stackrel{\varphi_a}{\leadsto} a$\,.  By the second equivalence in eq.(\ref{SasakiLeadsto}) we moreover obtain
\bcl 
$b\vdash(a\stackrel{S}{\to}c)\ \Longleftrightarrow\ b\stackrel{\varphi_a}{\leadsto}c$\,.
\ecl
The case of the backward relation is less straightforward (and in a sense also
less canonical).  Indeed, both in the definitions of $e^*$ and $e_*$ we use a forwardly expressed condition in terms of
``actuality at $t_1$ guarantees actuality at $t_2$''\,, where the
definition of
$\stackrel{e}{\looparrowleft}$ points backwardly\,.  However, we can quite easily prove a similar result as exposed in
eq.(\ref{SasakiLeadsto})\,.  We first do this for a general environment $e$\,.
\bpr\label{BackwCausRel}
$a_1\geq e_*(a_2)\ \Longleftrightarrow\ a_1\stackrel{e}{\looparrowleft}a_2$ and thus $a_1\geq\varphi_{a,*}(a_2)\
\Longleftrightarrow\ a_1\stackrel{\varphi_a}{\looparrowleft}a_2$\,. 
\epr
\bpf 
$(\Longleftarrow):$ From $p\models e_*(a_2)$\,, by definition of $e_*(a_2)$ as ``guarantees actuality of $a_2$ at
$t_2$'', it follows that $\tilde{e}^*(\{p\})\models a_2$\,, so by $a_1\stackrel{e}{\looparrowleft}a_2$ we then obtain
$p\models a_1$ and thus $a_1\geq e_*(a_2)$\,.
$(\Longrightarrow):$ We will first prove that $e_*(a_2)\stackrel{e}{\looparrowleft}a_2$\,, i.e. $(\forall p)_1(p\models e_*(a_2) \Leftarrow\tilde{e}^*(\{p\})\models a_2)$\,.  Once this is done,
$a_1\stackrel{e}{\looparrowleft}a_2$ given that $a_1\geq e_{*}(a_2)$ now follows straightforwardly.
Since $e_*(a_2)$ is the weakest property that guarantees actuality of $a_2$ at $t_2$ we clearly have $p\models
e_*(a_2)$ for all states $p$ at $t_1$ that guarantee actuality of $a_2$\,, i.e. $\tilde{e}^*(\{p\})\models a_2$\,, so
we do have 
$e_{*}(a_2)\stackrel{e}{\looparrowleft}a_2$\,, what completes this proof.
\bcl
$(a\stackrel{S}{\to}c)\vdash b\ \Longleftrightarrow\ b\stackrel{\varphi_a}{\looparrowleft}c$\,.
\ecl
\bcl
$b=(a\stackrel{S}{\to}c)\ \Longleftrightarrow\
b\stackrel{\varphi_a}{\leadsto}c\ \&\ b\stackrel{\varphi_a}{\looparrowleft}c$\,. 
\ecl
Note that as a part of the proof of Proposition \ref{BackwCausRel} we obtained
$e_{*}(a_2)\stackrel{e}{\looparrowleft}a_2$\,, and that by Proposition \ref{BackwCausRel} itself we obtain an
alternative way of defining causal assignment $e_*$\,, namely as: 
\par\smallskip\noindent 
ii'. `Causal assignment of properties' $e_*:L_2\to L_1$ that assigns to any property $a_2\in L_2$
the strongest property $e_*(a_2)\in L_1$ whose actuality is implied by $\Box$-actuality of $a_2$ at time $t_2$\,.
\par\smallskip\noindent
This alternative definition clearly exhibits in a more manifest way the {\sl backwardness} of causal assignment, and
consequently, of the action of the Sasaki hook. One easily verifies that {\it contra} eq.(\ref{SasakiLeadsto}) for
the case of
$\stackrel{e}{\leadsto}$ there is no obvious expression of $\stackrel{e}{\looparrowleft}$ in terms of $e^*$\,. The
naive idea one could have to propose $e^*(a_1)\geq a_2$ breaks down on the fact that this would imply $e_*\dashv
e^*$ what forces $e^*$ and
$e_*$ to be mutually inverse, something that in general (obviously) doesn't hold.  We will now proceed by
`extending' the relations $\stackrel{\varphi_a}{\leadsto}$ and $\stackrel{\varphi_a}{\looparrowleft}$ to operations.
 
\bigskip\noindent
{\bf 6. THE SASAKI HOOK WITHIN DYNAMIC OPERATIONAL QUANTUM LOGIC}       

\medskip\noindent 
First note that for general environments $e$ the relations $\stackrel{e}{\leadsto}\ \subseteq L_1\times L_2$ and
$\stackrel{e}{\looparrowleft}\
\subseteq L_1\times L_2$ easily extend to ${\rm DI}(L_1)\times{\rm DI}(L_2)$ by replacing ``($\Box$-)actuality of
...''  by ``truth with respect to ($\Box$-)actuality of a member of ...''\,, explicitly, 
\beqn\label{semdefcaqusrel1Bis}
A_1\stackrel{e}{\leadsto}A_2\ \Longleftrightarrow\ (\forall p)_1(p\models A_1 \Rightarrow
\tilde{e}^*(\{p\})\models A_2)\\ \label{semdefcaqusrel2Bis}
A_1\stackrel{e}{\looparrowleft} A_2 \ \Longleftrightarrow\ (\forall p)_1(p\models A_1 \Leftarrow
\tilde{e}^*(\{p\})\models A_2) 
\eeqn
where $\tilde{e}^*(\{p\})\models A_2$ now stands for $\forall q\in\tilde{e}^*(\{p\}):q\models A_2$\,, i.e. $\forall
q\in\tilde{e}^*(\{p\}), \exists a_2\in A_2:q\models a_2$\,, and where one verifies that 
$\downarrow\!a_1\stackrel{e}{\leadsto}\,\downarrow\!a_2\ \Longleftrightarrow\ 
a_1\stackrel{e}{\leadsto}a_2$ and $\downarrow\!a_1\stackrel{e}{\looparrowleft}\,\downarrow\!a_2\ \Longleftrightarrow\ 
a_1\stackrel{e}{\looparrowleft}a_2$\,.
Thus we can write elements of the image of ${\cal R}$, i.e., those elements in ${\rm
DI}(L)$ that represent properties, by the properties themselves.  
In view of eq.(\ref{eq:ImpVia Kripke})\,,
eq.(\ref{semdefcaqusrel1Bis}) and eq.(\ref{semdefcaqusrel2Bis}) it seems natural to set
\beqn\label{semdefcaqusIMP1Bis}
\mu(A_1\stackrel{e}{\to}A_2):=\{p\in\Sigma_1|p\models A_1 \Rightarrow \tilde{e}^*(\{p\})\models 
A_2\}\\   
\label{semdefcaqusIMP2Bis}
\mu(A_1\stackrel{e}{\leftarrow} A_2):=\{p\in\Sigma_1|p\models A_1 \Leftarrow
\tilde{e}^*(\{p\})\models A_2\}
\eeqn
indeed yielding an extension of the relations since
\beq
A_1\stackrel{e}{\leadsto}A_2\ \Longleftrightarrow\ (A_1\stackrel{e}{\to}A_2)=L 
\quad\quad{\rm and}\quad\quad
A_1\stackrel{e}{\looparrowleft}A_2\ \Longleftrightarrow\ (A_1\stackrel{e}{\leftarrow}A_2)=L\,. 
\eeq 
One verifies that on their turn
$(-\stackrel{e}{\to}-)$ and
$(-\stackrel{e}{\leftarrow}-)$ respectively define two tensors $(-\otimes_e-)$ and $(-\,_e\!\otimes-)$ via
adjunction (Coecke 2001b; Coecke, Moore and Smets 2001b; Smets 2001), i.e.
\beq
(A\otimes_e-)\dashv(A\stackrel{e}{\to}-)\quad{\rm and}\quad(-\,_e\!\otimes A)\dashv(-\stackrel{e}{\leftarrow}A)\,.
\eeq
In order to understand the significance of these tensors, first observe that the causal duality derived in the
previous section for maps respectively expressing propagation and causation for properties, can also be derived for
maps expressing propagation and causation of the propositions in ${\rm DI}(L)$\,, or equivalently\,, expressing
propagation and causation of sets of states in ${\rm F}(\Sigma)$\,.  This actually corresponds to forgetting about
the existence of ${\cal R}$ and applying the construction towards causal duality as if ${\rm DI}(L)\cong{\rm
F}(\Sigma)$ is the lattice of properties of a classical system.  The existence of a right adjoint of the map
$\hat{e}^*:{\rm DI}(L_1)\to{\rm DI}(L_2)$ that assigns to $A_1\in{\rm DI}(L_1)$ the strongest proposition in ${\rm
DI}(L_2)$ of which truth (i.e. actuality of a member) is implied by that of $A_1$\,, or equivalently, of the map
$\tilde{e}^*:{\rm F}(\Sigma_1)\to{\rm F}(\Sigma_2)$ that assigns to $T_1\in{\rm F}(\Sigma_1)$ the collection in
${\rm F}(\Sigma_2)$ of obtainable outcome states given that the initial state is in $T_1$\,, then imply preservation
of respectively $\bigvee_{{\rm DI}(L)}$ and $\bigcup$\,, i.e. disjunction. Complementary, since existence of a right
adjoint for propagation of properties encodes preservation of joins for properties, it follows from Coecke and
Stubbe (1999) and Coecke (2001a) that 
$\bigvee_L A=\bigvee_L B\ \Longrightarrow\ \bigvee_L\hat{e}^*(A)=\bigvee_L\hat{e}^*(B)$\,, 
or, expressed within ${\rm DI}(L)$\,, 
\beq\label{continuity2}
{\cal R}_{{\rm DI}(L)}(A)={\cal R}_{{\rm DI}(L)}(B)\ \ \Longrightarrow\ \
{\cal R}_{{\rm DI}(L)}(\hat{e}^*(A))={\cal R}_{{\rm DI}(L)}(\hat{e}^*(B))\,.      
\eeq
Thus, a shift ``from classical to quantum'' implies, besides the emergence of the operation ${\cal R}$\,, that
classical ``preservation of disjunction'' becomes a pair consisting of i. preservation of disjunction, and, ii. the
continuity-like condition of eq.(\ref{continuity2})\,.  Thus,
coexistence of laws on propagation at the level of $L$ and ${\rm DI}(L)$ is not a redundancy.   
\footnote{Note here that where $L$ induced a
closure on ${\rm DI}(L)$\,, we formally obtain in this case a restriction on the corresponding hom-sets ${\bf
SL}({\rm DI}(L_1),{\rm DI}(L_2))$ in ${\bf SL}$\,, the category of complete lattices and join-preserving maps. As
shown in Coecke and Stubbe (1999)\,, the physically admissible transitions constitute a subset of ${\bf SL}({\rm
DI}(L_1),{\rm DI}(L_2))$ for which there exists a `quantaloid morphism'
${\rm R}:{\bf SL}({\rm DI}(L_1),{\rm DI}(L_2))\to{\bf SL}(L_1,L_2)$\,.}
One now verifies that 
\beq 
(L\otimes_e-)=\hat{e}^*(-)\quad\quad\quad\quad{\rm and}\quad\quad\quad\quad(L\,_e\!\otimes-)=\hat{e}_*(-)\,,
\eeq
%%%%%
%%%%
%%% 
%% 
%
from which follow preservation properties with respect to meet and join, additionally to the ones that follow from the
fact that the tensors encode the left-adjoint actions to the hooks.\footnote{It turns out that $(-\otimes_e-)$ and
$(-\,_e\!\otimes-)$ respectively provide ${\rm DI}(L)$ with the structure of a commutative quantale and an in general
non-commutative dual quantale (Coecke 2001b; Coecke, Moore and Smets 2001b)\,.} 

\smallskip
%  
%%  
%%%
%%%%
%%%%%
How does all this apply to the context of quantum measurements, and as such, how does the Sasaki adjunction fits in
at this point.  First note that we have 
\beq
\hat{\varphi}^*_a:{\rm DI}(L_1)\to {\rm DI}(L_2): B\mapsto \bigvee_{{\rm DI}(L)}\{\,\downarrow\!\varphi^*_a(b)|b\in
B\}\ ;\,\downarrow\! b\mapsto\,\downarrow\!
\varphi^*_a(b)\,,
\eeq
i.e. $\varphi_{a}^*$ and $\hat{\varphi}_{a}^*$ act in the same on properties due to the eliminated emergence of
disjunction in $\varphi_{a}$\,. Do we have the same correspondens for the action of $\varphi_{a,*}$ and
$\hat{\varphi}_{a,*}$\,?
\bpr\label{DetAdj}
Given $f^*\dashv f_*:L_1\to L_2$ and $\hat{f}^*\dashv\hat{f}_*:{\rm DI}(L_1)\to {\rm DI}(L_2)$\,, then
$\,\downarrow\!f^*(-)=\hat{f}^*(\,\downarrow\!-)$ on $L_1$ implies $\,\downarrow\!f_*(-)=\hat{f}_*(\,\downarrow\!-)$
on $L_2$\,.
\epr
\bpf
For $a,b\in L :\hat{f}^*(\,\downarrow\! b)\subseteq\,\downarrow\! a\ \Longleftrightarrow
\,\downarrow\! \hat{f}^*(b)\subseteq\,\downarrow\! a\ \Longleftrightarrow
\hat{f}^*(b)\leq a\ \Longleftrightarrow b\leq f_*(a)$ so $\hat{f}^*(\,\downarrow\! a)=
\bigvee_{{\rm DI}(L_2)}\{B\in{\rm DI}(L_2)|\hat{f}^*(B)\subseteq\,\downarrow\! a\}=
\bigvee_{{\rm DI}(L_2)}\{\downarrow\! b|b\in L_2\,, b\leq f_*(a)\}=
\bigvee_{{\rm DI}(L_2)}\{\downarrow\!f_*(a)\}=\downarrow\!f_*(a)$\,.
\epf

\medskip\noindent
Thus, $\hat{\varphi}_{a,*}$ acts on properties as the Sasaki hook does, and it makes therefore
sense to set 
\beq
(a\stackrel{S}{\to}-):=\hat{\varphi}_{a,*}(-):{\rm DI}(L)\to{\rm DI}(L)\,.
\eeq
What do we obtain
in case for eq.(\ref{semdefcaqusIMP1Bis}) and eq.(\ref{semdefcaqusIMP2Bis}) in particular for properties as
arguments? Setting $^cT:=\Sigma\setminus T$ we obtain\,\footnote{E.g.  
via $\mu(a_1 \stackrel{\varphi_1}{\to} a_2)  =  \{p \in \Sigma_1 \mid p
\models a_1 \Rightarrow\varphi^{\ast}_{a}(p) \models a_2 \}
=\{p \in \Sigma_1 \mid p \models a_1 \Rightarrow p \models
\varphi_{\ast,a}(a_2)\}=\,^c\!\mu(a_1)\cup\mu(a \stackrel{S}{\to} a_2)$\,.}
\beq
\mu(a_1\stackrel{\varphi_a}{\to}a_2)=\,^c\!\mu(a_1)\cup\mu(a\stackrel{S}{\to}a_2)\quad\quad{\rm
and}\quad\quad
\mu(a_1\stackrel{\varphi_a}{\leftarrow}a_2)=\,^c\!\mu(a\stackrel{S}{\to}a_2)\cup\mu(a_1)\,.
\eeq
Note here in particular that the Sasaki hook does appear in the expression of both
$(a_1\stackrel{\varphi_a}{\to}a_2)$ and $(a_1\stackrel{\varphi_a}{\leftarrow}a_2)$\,, however with a different
antecedent than $(a_1\stackrel{\varphi_a}{\to}a_2)$ and $(a_1\stackrel{\varphi_a}{\leftarrow}a_2)$\,.
One verifies that using 
\beq\label{ExplNewHook}
(a_1\stackrel{\varphi_a}{\to}a_2)=(a_1\to_L \varphi_{a,*}(a_2))
\quad\quad\quad{\rm and}\quad\quad\quad
(a_1\stackrel{\varphi_a}{\leftarrow}a_2)=(\varphi_{a,*}(a_2)\to_L a_1)\,,
\eeq 
obtained via adjointness of $\varphi^*_a$ and $\varphi_{a,*}$ and Proposition \ref{DetAdj}\,,
and eq.(\ref{ImpForProp}) and eq.(\ref{PseudoAdj}) we obtain 
\beq\label{SasakiDynImp}
\begin{array}{r}
(a_1\stackrel{\varphi_a}{\to}a_2)=\{b\in L_1|a_1\wedge b\leq(a\stackrel{S}{\to}a_2)\}\\
=(a_1\to_L(a\stackrel{S}{\to} a_2))\hspace{16.7mm}  
\end{array}
\quad\ {\rm and}\quad \ 
\begin{array}{r}
(a_1\stackrel{\varphi_a}{\leftarrow}a_2)=\{b\in L_1|(a\stackrel{S}{\to}a_2)\wedge b\leq a_1\}\\
=((a\stackrel{S}{\to}a_2)\to_L a_1)\hspace{16.7mm}
\end{array}
\eeq
i.e.  respectively a forward and a backward dynamic $\varphi_a$-modification of the static hook $(-\to_L-)$\,.
Similar equations can be obtained for arguments in ${\rm DI}(L_1)\times{\rm DI}(L_2)$\,.
For the tensors we obtain  
\beq
(A_1\otimes_{\varphi_a}\!A_2)=\hat{\varphi}^*_a(A_1\wedge_{{\rm DI}(L)}A_2) 
\quad\quad\quad{\rm and}\quad\quad\quad
({A_1\,}_{\varphi_a}\!\!\otimes A_2)=(A_1\wedge_{{\rm DI}(L)}(a\stackrel{S}{\to}A_2))\,. 
\eeq 
%%% 
%% 
%
By construction we have the following modified versions of deduction and modus ponens (we express them for properties in
analogy to eq.(\ref{ForwHeytAdj}) and eq.(\ref{BackwHeytAdj})\,, the significance of the tensors is obvious):
\beq
b\otimes_{\varphi_a}\!c\vdash d\,\Rightarrow\,c\vdash (b\stackrel{\varphi_a}{\to}d)
\quad\ 
b\otimes_{\varphi_a}\!(b\stackrel{\varphi_a}{\to}c)\vdash c
\quad\
{b\,}_{\varphi_a}\!\!\otimes c\vdash d\,\Rightarrow\,b\vdash (d\stackrel{\varphi_a}{\leftarrow}c)
\quad\
{(c\stackrel{\varphi_a}{\leftarrow}b)\,}_{\varphi_a}\!\!\!\otimes b\vdash c\,. 
\eeq  
% 
%%  
%%%
We give a global overview of where the Sasaki operations fit    
in within DOQL --- we introduced the notations $\,\downarrow\![L]:=\{\downarrow\!a\,|\,a\in L\}$\,,
$(-\stackrel{S}{\leftarrow}a):=(a\stackrel{S}{\to}-)$ and
$(-\leftarrow_X\,\cdot\,):=(\,\cdot\,\to_X-)$ for $X\in\{L,{\rm DI}(L)\}$\,.\footnote{We note that there are some
subtilities which we didn't mention for sake of transparancy of the argument, in particular with respect to 
$\varphi_a^*(b)=0$ where we have two options: i. introduce a kernel of inadmissible initial states, i.e., consider
$\tilde{\varphi}_a^*:{\rm F}(\Sigma\setminus K)\to {\rm F}(\Sigma)$\,, or, consider upper pointed extensions
{\it sensu} (Coecke, Moore and Stubbe 2001; Sourbron 2001)\,. Note that similar considerations can be made for
environments $\varphi_{\{a,a'\}}$ instead of $\varphi_a$\, although then we truly obtain two levels, one for
$\varphi_{\{a,a'\}}^*$ and $\varphi_{\{a,a'\}_*}$ and one for
$\hat{\varphi}_{\{a,a'\}}^*$ and $\hat{\varphi}_{\{a,a'\}_*}$\,.   
}
 
\begin{center}       
\begin{tabular}{|c|c|c|}
\hline
{\bf Statical} (freezed dynamical) & {\bf $\varphi_a$-}induced forward {\bf dynamical} &  
{\bf $\varphi_a$-}induced backward {\bf dynamical} \\ 
\hline 
\hline 
$id_L:L\to L$ & 
$\varphi_a^*(-):L_1\to L_2$ & 
%%%%%
%%%%
%%% 
%% 
%
$\varphi_{a,*}(-)=(-\stackrel{S}{\leftarrow}a):L_2\to L_1$ \\  
% 
%%  
%%%
%%%%
%%%%%
\hline 
$id_{{\rm DI}(L)}:{\rm DI}(L)\to {\rm DI}(L)$ & 
$\hat{\varphi}_a^*(-)=\,\downarrow\!\varphi_a^*(-)\ on\, \downarrow\![L_1]$ & 
%%%%%
%%%%
%%% 
%% 
%
$\hat{\varphi}_{a,*}(-)=\,\downarrow\!(-\stackrel{S}{\leftarrow}a)\ on\, \downarrow\![L_2]$\\    
% 
%%  
%%%
%%%%
%%%%%
\hline
$(-\to_L-):L\times L\to{\rm DI}(L)$ &   
$(-\stackrel{\varphi_a}{\to}-)=(-\to_L(a\stackrel{S}{\to}-))$ & 
$(-\stackrel{\varphi_a}{\leftarrow}-)=(-\leftarrow_L(-\stackrel{S}{\leftarrow}a))$ \\
\hline
$(-\to_{{\rm DI}(L)}-)$\ on\, ${\rm DI}(L)$&  
$(-\stackrel{\varphi_a}{\to}-)=(-\to_{{\rm DI}(L)}(a\stackrel{S}{\to}-))$ &
$(-\stackrel{\varphi_a}{\leftarrow}-)=(-\leftarrow_{{\rm DI}(L)}(-\stackrel{S}{\leftarrow}a))$ \\
\hline
$(-\wedge_{{\rm DI}(L)}-)$\ on\, ${\rm DI}(L)$&
$(-\otimes_{\varphi_a}\!-)=\hat{\varphi}^*_a(-\wedge_{{\rm DI}(L)}-)$&
%%%%%
%%%%
%%% 
%% 
%
$(-_{\varphi_a}\!\!\otimes-)=(-\wedge_{{\rm DI}(L)}(-\stackrel{S}{\leftarrow}a))$\\
% 
%%  
%%%
%%%%
%%%%%
\hline
\end{tabular}
\end{center}
%%% 
%%  
%
Using eq.(\ref{ExplNewHook}) we can actually 
formally recover the Sasaki hook and projections as  
\beq
(1\stackrel{\varphi_a}{\to}-)=(1\to_L(a\stackrel{S}{\to}-))=(a\stackrel{S}{\to}-)\quad
\quad\quad\quad
%%%%%
%%%%
%%% 
%% 
% 
(1\otimes_{\varphi_a}\!-)=
{\varphi}^*_a(1\wedge-)=
{\varphi}^*_a(-)\,.
% 
%%  
%%% 
%%%%
%%%%%
\eeq
In the static case both of these become the identity, i.e. $(1{\to}_L-)=(1\wedge-)=id_L$\,, this giving the Sasaki hook
and projectors a formal interpretation as dynamic modifications of the identity.  Let us conclude this section with the
following identities
\beq
((a\stackrel{S}{\to}b)\stackrel{\varphi_a}{\to}b)=L\quad\quad\quad\quad
\quad\quad\quad\quad((a\stackrel{S}{\to}b)\stackrel{\varphi_a}{\leftarrow}b)=L
\eeq
or, differently put 
\beq
(a\stackrel{S}{\to}b)\stackrel{\varphi_a}{\leadsto}b\quad\quad\quad\quad
\quad\quad\quad\quad\quad\quad\quad(a\stackrel{S}{\to}b)\stackrel{\varphi_a}{\looparrowleft}b
\eeq
what formally encodes our interpretation of what the Sasaki hook does.
%  
%%   
%%%

\bigskip\noindent
{\bf 7. QUANTUM LOGIC RESEARCH? HOW TO CONVERT?}      

\medskip\noindent 
As mentioned above, Coecke (2001a), the traditional domain of study of quantum-like lattices should now be envisioned as a
study of the range of the operation ${\cal R}$\,, with corresponding heuristics.  This obviously sheds a different light
on the significance of, for example, orthomodularity.     
%%% 
%%  
%
Weak modularity actually doesn't come in at all in the disjunctive extension construction --- we recall from Coecke (2001a)
that orthocomplementation comes in in the sense that it means that operational resolution has an
involutive square-root named the `operational complementation'.  However, since orthomodularity is equivalent with the
Sasaki adjunction and this Sasaki adjunction in its turn represents causal duality, we have provided a new dynamic
interpretation of the axiom of orthomodularity. But, we can push this further.
% 
%%  
%%%
The essential
conclusion that comes out of our analysis is that in traditional quantum logic ``too much was encoded in too little'':
properties where identified with propositions, temporal phenomena like `change of state in a measurement' where statically
encoded, the distinction between the structure of properties and event structures was in many occasions mixed up, etc.  
What about event structures? Since we claimed that there was no conflict between an endo- and an exo-perspective, where
could they fit in in our setting?  They clearly should lie at the base of structuring the
collection of environments.\footnote{Some very general attempts in this direction were initiated in Amira,
Coecke and Stubbe (1998)\,.  We also mention that in the Foulis-Randall perspective there are some recent attempts by
Greechie and Gudder (2001) to build so-called sequential effect algebras that provide a dynamic structure for
discussing empirical events.  Along the lines of section 2 one could say that sequential effect algebras aim at
characterizing the structure of consecutive measurements as a reflection of the system's behavior there where the aim in
Amira, Coecke and Stubbe (1998) is essentially to obtain a theory on the system's behavior itself, incorporating the
interaction with its environment possibly including a measurement setup.} From a logical perspective, this implies a
two-dimensional situation: a structure of environments that interacts with a structure of propositions on a system.  Of
particular interest would then be the case  where we restrict to environments $\{\varphi_a|a\in L\}$\,, as we essentially
did in this paper.  This would mean that the environments are structured in the same way as the image of ${\cal R}$\,,
order-isomorphic to the properties
$L$\,, 
%%%  
%% 
%
so this realizes a structure in which the lattice of closed subspaces of a Hilbert space appears both as
the properties and as labels encoding (physical) environments, where actually we rather think about the latter as being
the projectors in standard
quantum theory.   In that sense the situation of ``too much being encoded in too little'' gets explicitly unraveled, and
motivates that it truly seems to make sense to distinguish between closed subspaces and projectors at an
abstract structural level although they are in bijective correspondence, the first having an ontological connotation, the
second an empirical. This then leads to the  
perspective that the transition from either classical or 
constructive/intuitionistic logic to quantum logic entails besides the introduction of an
additional unary connective operational resolution 
%%%%%
%%%%
%%% 
%% 
%
${\cal R}:{\rm DI}(L)\to{\rm DI}(L)$    
% 
%%  
%%%
%%%%
%%%%%
the shift from a binary connective
implication to a ternary connective 
\beq
%%%%%
%%%%
%%% 
%% 
%
(- \stackrel{-}{\to} -):{\rm DI}(L)\times {\cal R}[{\rm DI}(L)]\times {\rm DI}(L)\to {\rm DI}(L)\,.  
% 
%%  
%%%
%%%%
%%%%%
\eeq
where two of the arguments refer to qualities of the system and the third, the new one, to
an obtained outcome (in a measurement).  A better way of putting things would be 
\beq 
%%%%%
%%%%
%%% 
%% 
%
(- \stackrel{-}{\to} -):{\rm DI}(L)\times {\Bbb P}(L)\times {\rm DI}(L)\to {\rm DI}(L)\,.  
% 
%%  
%%% 
%%%%
%%%%%
\eeq
where ${\Bbb P}(L)$ are the projectors (on corresponding closed subspaces) what for general orthomodular
lattices ends up being the Baer $^*$-semigroup of projectors in the Foulis (1960) sense. In view of Piron's (1964)
theorem it then follows that this situation fully covers `pointless quantum theory' --- sensu pointless topology in terms
of locales, i.e. complete Heyting algebras --- since we drop the two point-related axioms atomisticity and covering law in
order to have a (complete) orthomodular lattice.
Another extrapolation of the setting presented in this paper consists of rather than defining dynamic operations (labeled
by environments) on the static propositions
${\rm DI}(L)$\,, we start of from `dynamic propositions', i.e. propositions on `propagation of (actual) properties' rather
than on actuality itself and here inspiration can be found in research within the domain of for example computational
process semantics (Milner 1999), action logic (Baltag 1999) etc. Obviously, much is still to be done in that direction.  
% 
%%  
%%%

\bigskip\noindent
{\bf 8. ACKNOWLEDGEMENTS}

\medskip\noindent 
We thank Samson Abramsky, Marisa Dalla-Chiara, David Foulis, Dick Greechie, Chris Isham, Jim Lambek, David Moore, Ioannis
Raptis and Isar Stubbe for  discussions and comments that have
led to the present content and form of the presentation in this paper. Part of the research reflected in this paper was performed by Bob Coecke at
{\sl McGill University, Department of  Mathematics and Statistics, Montreal} and {\sl Imperial College of Science,
Technology \& Medicine, Theoretical Physics Group, London}\,.  Bob Coecke is Postdoctoral Researcher at the 
European TMR Network ``Linear Logic in Computer Science''. Sonja Smets is Postdoctoral Researcher at
Flanders' Fund for Scientific Research.    

\end{document}